\documentclass[twocolumn]{aastex63}

\usepackage[utf8]{inputenc} % allow utf-8 input
\usepackage[T1]{fontenc}    % use 8-bit T1 fonts
\usepackage{hyperref}       % hyperlinks
\usepackage{url}            % simple URL typesetting
\usepackage{booktabs}       % professional-quality tables
\usepackage{amsfonts}       % blackboard math symbols
\usepackage{nicefrac}       % compact symbols for 1/2, etc.
\usepackage{microtype}      % microtypography
\usepackage{amsmath}
\usepackage{ragged2e}
\usepackage{float}
\usepackage{graphicx}
\usepackage{natbib}
\usepackage[english]{babel}
\usepackage[normalem]{ulem}
\usepackage{cancel}

\graphicspath{{./}{figures/}}
\begin{document}

\title{A Quasi-Linear Diffusion Model for Resonant Wave-Particle Instability in Homogeneous Plasma}

\author{Seong-Yeop Jeong}
\affiliation{Mullard Space Science Laboratory, University College London, Dorking, RH5 6NT, UK; s.jeong.17@ucl.ac.uk, d.verscharen@ucl.ac.uk}

\author{Daniel Verscharen}
\affiliation{Mullard Space Science Laboratory, University College London, Dorking, RH5 6NT, UK; s.jeong.17@ucl.ac.uk, d.verscharen@ucl.ac.uk}
\affiliation{Space Science Center, University of New Hampshire, Durham, NH 03824, USA}

\author{Robert T. Wicks}
\affiliation{Mullard Space Science Laboratory, University College London, Dorking, RH5 6NT, UK; s.jeong.17@ucl.ac.uk, d.verscharen@ucl.ac.uk}
\affiliation{Institute for Risk and Disaster Reduction, University College London, Gower Street, London, WC1E 6BT, UK}

\author{Andrew N. Fazakerley}
\affiliation{Mullard Space Science Laboratory, University College London, Dorking, RH5 6NT, UK; s.jeong.17@ucl.ac.uk, d.verscharen@ucl.ac.uk}



\begin{abstract}

In this paper, we develop a model to describe the generalized wave-particle instability in a quasi-neutral plasma. We analyze the quasi-linear diffusion equation for particles by expressing an arbitrary unstable and resonant wave mode as a Gaussian wave packet, allowing for an arbitrary direction of propagation with respect to the background magnetic field. We show that the localized energy density of the Gaussian wave packet determines the velocity-space range in which the dominant wave-particle instability and counter-acting damping contributions are effective. Moreover, we derive a relation describing the diffusive trajectories of resonant particles in velocity space under the action of such an interplay between the wave-particle instability and damping. For the numerical computation of our theoretical model, we develop a mathematical approach based on the Crank-Nicolson scheme to solve the full quasi-linear diffusion equation. Our numerical analysis solves the time evolution of the velocity distribution function under the action of a dominant wave-particle instability and counteracting damping and shows a good agreement with our theoretical description. As an application, we use our model to study the oblique fast-magnetosonic/whistler instability, which is proposed as a scattering mechanism for strahl electrons in the solar wind. In addition, we numerically solve the full Fokker-Planck equation to compute the time evolution of the electron-strahl distribution function under the action of Coulomb collisions with core electrons and protons after the collisionless action of the oblique fast-magnetosonic/whistler instability. %\textbf{We compare the timescales between strahl-scattering and collisional relaxation. We find that under typical solar-wind conditions at 0.3au from the Sun, the collisional timescale is greater by a factor of $10^{5}$ than the scattering timescale.}

\end{abstract}



\section{Introduction} \label{intro}

%The wave-particle resonance is a phenomenon where the charged particles in a plasma have a specific velocity, so that the particles itself experience the wave as a static electric field.

Wave-particle resonances play an important role for the energy exchange between particles and waves in many space and astrophysical plasmas. For example, wave-particle resonances contribute to the acceleration and deceleration of particles in radiation belts \citep{Ukhorskiy2014}, the deviation of the particle velocity distribution function (VDF) from a Maxwellian equilibrium in the solar wind \citep{Marsch2006}, the thermodynamic state of the intracluster medium in galaxy clusters \citep{Roberg_Clark_2016}, and the scattering and absorption of the surface radiation in neutron-star magnetospheres \citep{10.1111/j.1365-2966.2006.10140.x}. Therefore, it is of great importance to study the mechanics of wave-particle resonances in order to advance our understanding of the physics of astrophysical plasmas throughout the universe. 

According to kinetic theory, wave-particle resonances can occur in the form of Landau or cyclotron resonances, which contribute to wave instability or wave damping depending on the resonance's characteristics. The quasi-linear theory of magnetized plasma, first established by \citet{osti_4703238} and \citet{1966PhFl....9.2377K}, provides a mathematical framework to predict the evolution of the particle VDF under the action of wave-particle resonances. Quasi-linear theory assumes that the spatially averaged VDF evolves slowly compared to the gyroperiod of the particles and the wave period. It furthermore assumes that the fluctuation amplitude is small and that the spatial average of the fluctuations vanishes. Based on this theory, numerous analytical studies have successfully explained the evolution of VDFs resulting from wave-particle resonances. 

Resonant particles diffuse along specific trajectories in velocity space determined by the properties of resonant wave
\citep{1966PhFl....9.2377K,1968JATP...30.1313G,doi:10.1029/RG019i001p00171, doi:10.1029/JA085iA09p04577,1992wapl.book.....S,doi:10.1029/96JA00293,doi:10.1029/98JA01740,doi:10.1029/2000JA000309}.
In these models, quasi-linear diffusion coefficients determine the diffusion rate of the resonant particles \citep{lyons_thorne_kennel_1971, lyons_1974,doi:10.1029/2004SW000069,doi:10.1029/2004JA010851,doi:10.1029/2005JA011159,article2,article}. Alternatively, quasi-linear diffusion models based on a bi-Maxwellian VDF, in which only its moments evolve in time, describe the effective evolution of particle VDFs under the action of microinstabilities \citep{doi:10.1029/2012JA017645,doi:10.1029/2012JA017697,doi:10.1002/2015JA021495,doi:10.1063/1.4997666,2017RvMPP...1....4Y}. Moreover, fully nonlinear simulations based on kinetic theory model the evolution of the particle VDF consistently with predictions from quasi-linear theory \citep{article3,Vocks_2005,doi:10.1029/2007GL032327, doi:10.1063/1.2997339,doi:10.1063/1.3526602,doi:10.1063/1.3676155}. Observations from Helios revealed signatures in the proton VDFs consistent with ion cyclotron resonances as predicted by quasi-linear theory \citep{doi:10.1029/2000JA000414,doi:10.1029/2001JA000150,doi:10.1029/2006JA011979,angeo-29-2089-2011}.

Realistic analytical models must describe the diffusive trajectory of the resonant particles in velocity space, taking into account the localized (in wavevector space) energy density of the waves that resonate with these particles. These models must also account for non-Maxwellian features in the VDF evolution in order to advance our understanding of plasma observations and kinetic simulation results. A rigorous numerical analysis of the diffusion equation, including both the diagonal and off-diagonal diffusion terms, is necessary to support the theoretical description through the quantification of the diffusion rates.

By analyzing the quasi-linear diffusion equation, we propose a novel quasi-linear diffusion model for the time evolution of VDFs under the action of a dominant wave-particle instability and counter-acting damping contributions in Section~\ref{2}. Our model describes the creation and evolution of non-Maxwellian features in the particle VDF. We allow for an arbitrary type of the unstable and resonant wave mode with an arbitrary direction of propagation with respect to the background magnetic field. In our analysis, we express the electric field of this wave as a Gaussian wave packet in configuration space. The localization of such a wave packet in configuration space is the direct consequence of its generation through a linear instability, which is localized in wavevector space. 

To investigate the stabilization of the VDF through quasi-linear diffusion, we apply our analysis of the quasi-linear diffusion equation to Boltzmann's $H$-theorem. In this scheme, the localized energy density of the Gaussian wave packet in wavevector space defines the velocity-space range in which the dominant wave-particle instability and counter-acting damping contributions are effective. In addition, we derive a relation to describe the diffusive trajectories of resonant particles in velocity space under the action of such an instability and damping. In this way, our model accounts for the diffusive behavior of resonant particles in different regions of velocity space.

For the numerical evaluation of our theoretical description, we develop a mathematical approach based on the Crank-Nicolson scheme (for numerical details, see Appendix~\ref{A}) that solves the full quasi-linear diffusion equation. Because of its reliable stability, the Crank-Nicolson scheme has been used previously to solve diffusion equations in a variety of fields \citep{doi:10.1029/2001JA000180,doi:10.1029/2004SW000069,Br_gmann_2004,YANG200931,2016ApJ...820...47K,2019ApJ...882..157T}. However, most standard Crank-Nicolson schemes ignore the off-diagonal terms in the diffusion equation. In our case, these off-diagonal terms are important for the description of resonant pitch-angle scattering.
We note that our mathematical approach is applicable to all general two-dimensional diffusion equations, including those with off-diagonal diffusion terms.

In Section~\ref{3}, as an example, we apply our model to the scattering of the electron strahl, which is a field-aligned electron beam population in the solar wind \citep{doi:10.1029/JA092iA02p01075,doi:10.1029/2008JA013883}. Observations in the solar wind suggest that strahl electrons exchange energy with whistler waves, which ultimately leads to a scattering of strahl electrons into the halo population \citep{doi:10.1029/2006JA011967,2014ApJ...796....5L,doi:10.1002/2016JA023656}.
Our quasi-linear framework confirms that an instability of the fast-magnetosonic/whistler wave in oblique propagation with respect to the background magnetic field scatters the electron strahl into the electron halo, as predicted by linear theory \citep{2019ApJ...871L..29V,Verscharen_2019}. 

In Section~\ref{4}, for a more realistic model of the strahl evolution after the collisionless action of the oblique fast-magnetosonic/whistler instability, we numerically solve the full Fokker-Planck equation for Coulomb collisions with our mathematical approach (for numerical details, see Appendix~\ref{B}). We model the time evolution of the electron-strahl VDF through the action of Coulomb collisions with core electrons and protons. This combined method allows us to compare the timescales for the strahl scattering and collisional relaxation. 

In Section~\ref{5}, we discuss the results of our model for the strahl scattering and electron-halo formation through the instability and Coulomb collisions. In Section~\ref{6}, we summarize and conclude our treatment.

\section{Quasi-Linear Diffusion Model} \label{2}
In this section, we establish our general theoretical framework for the description of a resonant wave-particle instability in quasi-linear theory. Because our work focuses on non-relativistic space plasma like the solar wind, we ignore relativistic effects throughout our study.
\clearpage
\subsection{Analysis of the Quasi-Linear Diffusion Equation}\label{2.1}
To investigate the time evolution of the particle VDF through wave-particle resonances, we study the quasi-linear diffusion equation, given by \citet{1992wapl.book.....S}:
\begin{equation}\label{QLD}
\begin{split}
&\left (\frac{\partial  f_{j} }{\partial  t}\right)_{QLD}=\lim_{V\rightarrow \infty } \sum_{n=-\infty }^{\infty } \int  \frac{\pi q_j^{2}}{V m_j^2}  \\
&\times \! \hat{G}[k_\parallel] \!\Bigg[ \frac{v_{\perp}^2}{\left |v_{\parallel} \right |} \delta  \bigg ( k_\parallel\! -\! \frac{\omega_{k}\!-\!n\Omega_{j}}{v_{\parallel}}\bigg ) \!\! \left | \psi_j^{n} \right |^{2} \! \hat{G}[k_\parallel]f_{j} \! \Bigg ]  d^3\mathbf{k},
\end{split}
\end{equation}
where
\begin{equation}\label{psy}
\begin{split}
\psi_j^{n} 
\!\equiv\!\frac{1}{\sqrt{2}}  \bigg [ \mathit{{E}}_{k}^{R} e^{i\phi}  J_{n+1}(\rho_j) +\mathit{{E}}_{k}^{L}& e^{-i\phi} J_{n-1}(\rho_j) \bigg ]\\
&+\!\frac{v_\parallel}{v_\perp} \mathit{{E}}_{k}^{z}J_{n}(\rho_j),
\end{split}
\end{equation}
and
\begin{equation}\label{g}
\hat{G}[k_\parallel] \equiv \left ( 1 - \frac{ k_\parallel v_\parallel }{\omega_{k}} \right ) \frac{1}{v_\perp} \frac{\partial }{\partial v_\perp}+ \frac{ k_\parallel }{\omega_{k}} \frac{\partial }{\partial v_\parallel}.
\end{equation}
The integer $n$ determines the order of the resonance, where $n=0$ corresponds to the Landau resonance and $n\neq 0$ corresponds to cyclotron resonances. In our equations, we label contributions from a given resonance order with a  superscript $n$. The subscript \textit{j} indicates the particle species. The particle VDF of species \textit{j} is denoted as $f_{j}\equiv f_{j}(v_\perp,v_\parallel,t)$ which is spatially averaged and gyrotropic, $q_j$ and $m_j$ are the charge and mass of a particle of species \textit{j}, $v_\perp$ and $v_\parallel$ are the velocity coordinates perpendicular and parallel with respect to the background magnetic field. We choose the coordinate system in which the proton bulk velocity is zero. %We adopt the definition of the delta function as the limit of the sequence of Gaussians in our analysis:
%\begin{equation}\label{delta}
%\delta(x) \equiv  \lim_{a\rightarrow 0 }  \frac{1}{\left |a \right |\sqrt{\pi}}\exp \left[-\frac{x^2}{a^2}\right].
%\end{equation}
We denote the $n$th-order Bessel function as $J_n(\rho_j)$ where $\rho_j \equiv k_\perp v_\perp/\Omega_{j}$.  The cyclotron frequency of species \textit{j} is defined as $\Omega_j \equiv q_j B_0/m_jc$, $\mathbf B_0$ is the background magnetic field, $c$ is the speed of light, $k_\perp$ and $k_\parallel$ are the perpendicular and parallel components of the wavevector $\mathbf{k}$, and $V$ is the volume in which the wave amplitude is effective so that the wave and particles undergo a significant interaction. We denote Dirac's $\delta$-function as $\delta$ and the azimuthal angle of wavevector $\mathbf{k}$ by $\phi$. The frequency $\omega$ is a complex function of $\mathbf{k}$, and we define $\omega_{k}$ as its real part and $\gamma_k$ as its imaginary part ($\omega=\omega_{k}+i\gamma_k$). Without loss of generality, we set $\omega_{k}>0$. Furthermore, we assume that $|\gamma_k|\ll \omega_{k}$, i.e. the assumption of slow growth or damping that is central to quasi-linear theory.

The spatially Fourier-transformed electric field has the form of $\mathbf{E}_{\mathbf k}=\hat{x}E_k^x+\hat{y}E_k^y+\hat{z}E_k^z$ and is defined as \citep{gurnett_bhattacharjee_2017}
\begin{equation}\label{ft}
    \mathbf{E}_{\mathbf k}\!=\!\frac{1}{(2\pi)^{3/2}} \int \mathbf{E}_{\mathbf r} \exp \left[-i\mathbf k \cdot \mathbf r \right] d^3\mathbf{r},
\end{equation}
where $\mathbf k \cdot \mathbf r=k_{\perp}x\cos\phi+k_{\perp}y\sin\phi+k_{\parallel}z$ and $\mathbf{r}$ is the position vector. We take the constant background magnetic field as $\mathbf{B}_0=\hat z B_0$ and define the right- and left-circularly polarized components of the electric field as $\mathit{{E}}_{k}^{R}\equiv(\mathit{{E}}_{k}^{x}-i\mathit{{E}}_{k}^{y})/\sqrt{2}$ and $\mathit{{E}}_{k}^{L}\equiv(\mathit{{E}}_{k}^{x}+i\mathit{{E}}_{k}^{y})/\sqrt{2}$.
The longitudinal component of the electric field is $\mathit{{E}}_{k}^{z}$.

Linear instabilities typically create fluctuations across a finite range of wavevectors. The Fourier transformation of such a wave packet in wavevector space corresponds to a wave packet in configuration space. For the sake of simplicity, we model this finite wave packet by assuming that the electric field $\mathbf{E}_\mathbf{r}$ of the unstable and resonant waves has the shape of a gyrotropic Gaussian wave packet
\begin{equation}\label{e}
\begin{split}
    \!\!\!\mathbf{E}_{\mathbf r}=\mathbf{E}_{0}
    \exp \! \left [-\frac{\sigma_{\perp 0}^2 x^2\!+\!\sigma_{\perp 0}^2 y^2\!+\!\sigma_{\parallel 0}^2 z^2}{2} \right] \!\exp  \left[i\mathbf k_0 \cdot \mathbf r \right],
\end{split}
\end{equation}
where $\mathbf{E}_{0}= \hat x E_{0}^x+\hat y E_{0}^y+\hat z E_{0}^z$, $\mathbf k_0 \cdot \mathbf r=k_{\perp 0}x\cos\phi+k_{\perp 0}y\sin\phi+k_{\parallel 0}z$, and $\mathbf{k}_0$ is the wavevector of the Gaussian wave packet. We allow for an arbitrary angle $\theta_0$ between $\mathbf{k}_0$ and $\mathbf{B}_0$, which defines the orientation of the wavevector at maximum growth of the wave, and assume that $k_{\parallel 0}\neq 0$. The vector $\mathbf{E}_{0}$ represents the peak amplitude of the electric field. The free parameters $\sigma_{\perp0}$ and $\sigma_{\parallel0}$ characterize the width of the Gaussian envelope. Quasi-linear theory requires that $\mathbf{E}_{\mathbf r}$ spatially averages to zero. Therefore, we assume that $|k_{\parallel 0}|\gg\sigma_{\parallel0}$ so that the spatial dimension of the Gaussian wave packet is large compared to the parallel wavelength $2\pi/|k_{\parallel0}|$.

The spatial Fourier transformation of Eq.~(\ref{e}) according to Eq.~(\ref{ft}) then leads to
\begin{equation}\label{eft}
\begin{split}
\mathbf{{E}}_{\mathbf k}=& \frac{ \mathbf{E}_{\mathbf 0}}{ \sigma_{\parallel0} \sigma_{\perp0}^2}\exp \Bigg [-\frac{(k_\parallel-k_{\parallel0})^2}{2\sigma_{\parallel0}^2}-\frac{(k_\perp-k_{\perp0})^2}{2\sigma_{\perp0}^2} \Bigg].%(2\pi)^{3/2}
\end{split}
\end{equation}
We identify $V$ with the volume of the Gaussian envelope, $V=1/(\sigma_{\parallel0} \sigma_{\perp0}^2)$. Eq.~(\ref{eft})  describes the localization of the wave energy density in wavevector space. For the instability analysis through Eq.~(\ref{eft}), we define the unstable $\mathbf{k}$-spectrum as the finite wavevector range in which $\gamma_{k}>0$ and argue that resonant waves exist only in this unstable $\mathbf{k}$-spectrum. We ignore any waves outside this $\mathbf{k}$-spectrum since they are damped. 

We define $k_{\parallel 0}$ as the value of $k_\parallel$ at the center of the unstable $\mathbf{k}$ spectrum. We then obtain
\begin{equation}\label{perk}
    k_{\perp 0}=k_{\parallel 0}\tan\theta_0.
\end{equation}
In the case of a linear plasma instability, we identify $k_{\perp 0}$ and $k_{\parallel 0}$ with the wavevector components at which the instability has its maximum growth rate as a reasonable approximation. To approximate the wave frequency of the unstable waves at the angle $\theta_0$ of maximum growth, we expand $\omega_k$ of the unstable and resonant waves around $k_{\parallel 0}$ as
\begin{equation}\label{w}
    \omega_{k}(k_\parallel) \approx \omega_{k0}+v_{g0}\left ( k_\parallel-k_{\parallel 0} \right ),
\end{equation}
where
\begin{equation}\label{group}
v_{g0} \equiv \frac{\partial \omega_{k}}{\partial k_\parallel}\bigg|_{k_\parallel=k_{\parallel 0}}.
\end{equation}
In Eqs.~(\ref{w}) and (\ref{group}), $\omega_{k0}$ and $v_{g0}$ are the wave frequency and parallel group velocity of the unstable and resonant waves, evaluated at $k_\parallel=k_{\parallel 0}$. We select the values of $\sigma_{\perp0}$ and $\sigma_{\parallel0}$ as the half widths of the perpendicular and parallel unstable $\mathbf{k}$-spectrum. In the case of a linear plasma instability, the numerical values for $k_{\perp 0}$, $k_{\parallel 0}$, $\sigma_{\perp 0}$, $\sigma_{\parallel 0}$, $\omega_{k 0}$ and $v_{g0}$ can be found from the solutions of the hot-plasma dispersion relation, which thus closes our set of equations.

By using Eq.~(\ref{eft}) and Eq.~(\ref{w}), we rewrite Eq.~(\ref{QLD}) as
\begin{equation}\label{QLDQ}
    \left (\frac{\partial  f_{j} }{\partial  t}\right)_{QLD}=\sum_{n=-\infty }^{\infty } \int \hat{G}[k_{\parallel}]\big[D_j^{n}\hat{G}[k_{\parallel}] f_j\big]  d^3\mathbf{k},
\end{equation}%we define the diffusion coefficient for the $n$ resonance as
%and resolving the $\delta$-function pertaining to the $k_\parallel$-integral in Eq.~(\ref{QLD})
%\begin{equation}\label{D}
%\begin{split}
%D_j^{n}(v_{\perp},v_{\parallel})= \frac{q_j^2 v_{\perp}^2 W_j^{n}}{8 \pi^2\sigma_\parallel \sigma_\perp^2 m_j^2} \int_{0}^{2\pi} &\int_{0}^{\infty}  | \widetilde\psi_{j}^{n} |^{2} \\
%&\!\!\!\!\!\!\!\!\!\!\!\!\!\!\!\!\!\!\!\!\!\!\!\!\!\!\!\!\!\!\!\!\!\!\!\!\!\!\!\!\!\!\!\!\!\!\!\!\!\!\times\exp \Bigg [-\frac{(k_\perp-k_{\perp0})^2}{\sigma_\perp^2} \Bigg]  k_\perp  dk_\perp d\phi,
%\end{split}
%\end{equation}
where
\begin{equation}\label{iD}
\begin{split}%8 \pi^2
&D_j^{n} \equiv \frac{\pi q_j^2 v_{\perp}^2 }{\sigma_{\parallel0} \sigma_{\perp0}^2 m_j^2} \delta  (  k_\parallel  - k_{\parallel j}^{n}  ) \\
&\times \frac{| \psi_{j0}^{n} |^{2}}{|v_{\parallel}-v_{g0} |} \exp  \Bigg [-\frac{(k_\parallel\!-\!k_{\parallel0})^2}{\sigma_{\parallel0}^2}-\frac{(k_\perp\!-\!k_{\perp0})^2}{\sigma_{\perp0}^2} \Bigg],
\end{split}
\end{equation}
\begin{equation}\label{psyp}
\begin{split}
    \psi_{j0}^{n}\!\equiv\!\frac{1}{\sqrt{2}}   \bigg [ \mathit{{E}}_0^{R} e^{i\phi} J_{n+1}(\rho_j) +\mathit{{E}}_0^{L}& e^{-i\phi} J_{n-1}(\rho_j)  \bigg ] \\
    &+\frac{v_\parallel}{v_\perp} \mathit{{E}}_{0}^z J_{n}(\rho_j),
\end{split}
\end{equation}
and
\begin{equation}\label{res}
    k_{\parallel j}^{n}\equiv\frac{\omega_{k0}- k_{\parallel 0}v_{g0}-n\Omega_{j}}{v_{\parallel}-v_{g0}}.
\end{equation}
We set $\mathit{{E}}_0^{R}=(\mathit{{E}}_{0}^x-i\mathit{{E}}_{0}^y)/\sqrt{2}$ and $\mathit{{E}}_0^{L}=(\mathit{{E}}_{0}^x+i\mathit{{E}}_{0}^y)/\sqrt{2}$ as constant, evaluated at $\mathbf{k}_{0}$.
%\begin{equation}\label{QLDQ}
 %   \left (\frac{\partial  f_{j} }{\partial  t}\right)_{QLD}=\sum_{n=-\infty }^{\infty } \hat{G}[k_{\parallel j}^{n}]\Big[D_j^{n}\hat{G}[k_{\parallel j}^{n}] f_j\Big],
%\end{equation}

Eq.~(\ref{QLDQ}) is the quasi-linear diffusion equation describing the action of the dominant wave-particle instability and co-existing damping contributions from other resonances in a Gaussian wave packet. We define the $n$ resonance as the contribution to the summation in Eq.~(\ref{QLDQ}) with only integer $n$. We note that any $n$ resonance can contribute to wave instability or to wave damping depending on the resonance's characteristics. 

\subsection{Stabilization through a Resonant Wave-Particle Instability }\label{2.2}
We define stabilization as the process that creates the condition in which $(\partial f_j/\partial t)_{QLD}\rightarrow 0$ for all $v_{\perp}$ and $v_{\parallel}$. For our analysis of the stabilization of a VDF through a resonant wave-particle instability, including co-existing damping effects, we use Boltzmann's $H$-theorem, in which the quantity $H$ is defined as
\begin{equation}\label{H}
H(t)\equiv\int f_j(\mathbf{v},t) \ln f_j(\mathbf{v},t) d^3\mathbf{v}.
\end{equation}
By using Eq.~(\ref{QLDQ}), the time derivative of $H$ is given by
%\begin{equation}\label{Ht}
%\begin{split}
%\frac{d H}{d t}\!=\!\sum_{n=-\infty }^{\infty }\int\!(\ln f_j\!+\!1)\hat{G}[k_{\parallel j}^{n}]\Big[D_j^{n}\hat{G}[k_{\parallel j}^{n}]f_j\Big] d^3\mathbf{v}.
%\end{split}
%\end{equation}
\begin{equation}\label{Ht}
\begin{split}
\!\!\!\!\frac{d H}{d t}\!=\!\!\sum_{n=-\infty }^{\infty }\!\int\!\!\!\int\!(\ln f_j\!+\!1)\hat{G}[k_{\parallel}]\big[D_j^{n}\hat{G}[k_{\parallel}]f_j\big] d^3\mathbf{k}d^3\mathbf{v}.
\end{split}
\end{equation}
The integrand in Eq.~(\ref{Ht}) is equivalent to 
%\begin{equation}\label{int}
%\begin{split}
%( &\ln f_j+1)\hat{G}[k_{\parallel j}^{n}]\Big[D_j^{n}\hat{G}[k_{\parallel j}^{n}]f_j\Big]=\\
%&\hat{G}\![k_{\parallel j}^{n}]\Big[D_j^{n}\hat{G}[k_{\parallel j}^{n}](f_j \ln f_j)\Big]\!- D_j^{n}f_j^{-1}\Big[\hat{G}[k_{\parallel j}^{n}]f_j\Big]^{\!2}.
%\end{split}
%\end{equation}
\begin{equation}\label{int}
\begin{split}
( &\ln f_j+1)\hat{G}[k_{\parallel}]\big[D_j^{n}\hat{G}[k_{\parallel}]f_j\big]=\\
&\hat{G}[k_{\parallel}]\big[D_j^{n}\hat{G}[k_{\parallel}](f_j \ln f_j)\big]- D_j^{n}\big[\hat{G}[k_{\parallel}]f_j\big]^{2}\big / f_j.
\end{split}
\end{equation}
Upon substituting Eq.~(\ref{int}) into Eq.~(\ref{Ht}), the first term on the right-hand side in Eq.~(\ref{int}) disappears after the integration over $\mathbf v$. Then, by resolving the $\delta$-function in $D_j^{n}$ through the $k_\parallel$-integral, we obtain
\begin{equation}\label{Htt}
\frac{d H}{d t}=-\sum_{n=-\infty }^{\infty } \left(\frac{d H}{d t}\right )^n,
\end{equation}
where
\begin{equation}\label{Hn}
\begin{split}%8 \pi^2
\left(\frac{d H}{d t}\right )^n \equiv \int \Big\{ \widetilde D_j^{n} \big[\hat{G}[k_{\parallel j}^{n}]f_j\big]^2 \big / f_j \Big\} d^3\mathbf{v},
\end{split}
\end{equation}
\begin{equation}\label{D}
\begin{split}%8 \pi^2
\widetilde D_j^{n} \equiv & W_j^{n} \frac{\pi q_j^2 v_{\perp}^2}{\sigma_{\parallel0} \sigma_{\perp0}^2 m_j^2} \int_{0}^{2\pi} \int_{0}^{\infty}  | \psi_{j0}^{n} |^{2} \\
&\times\exp \Bigg [-\frac{(k_\perp-k_{\perp0})^2}{\sigma_{\perp0}^2} \Bigg]  k_\perp  dk_\perp d\phi,
\end{split}
\end{equation}
\begin{equation}\label{W}
    W_j^{n}\equiv\frac{1}{ |v_{\parallel}-v_{g0} |}\exp \left[ -\frac{k_{\parallel 0}^2}{\sigma_{\parallel0}^2}\left(\frac{v_{\parallel}-v_{\parallel res}^n}{v_{\parallel}-v_{g0}}\right)^{2} \right ],
\end{equation}
\begin{equation}\label{resvelo}
v_{\parallel res}^n\equiv\frac{\omega_{k0}-n\Omega_{j}}{k_{\parallel0}},
\end{equation}
\begin{equation}\label{gres}
\begin{split}
\hat{G}[k_{\parallel j}^{n}]\equiv\bigg[\frac{n\Omega_{j}}{\omega_{k0}- k_{\parallel 0}v_{g0}-n\Omega_{j}}\bigg ] \frac{v_\parallel-v_{g0}}{v_{ph}v_\perp}&\frac{\partial }{\partial v_\perp}\\
&\!\!\!\!\!\!\!\!\!\!\!\!\!\!\!+\frac{1}{v_{ph}} \frac{\partial }{\partial v_\parallel},
\end{split}
\end{equation}
and
\begin{equation}\label{respv}
    v_{ph}\equiv \frac{\omega_{k}(k_{\parallel j}^{n})}{k_{\parallel j}^{n}}=\frac{(\omega_{k0}- k_{\parallel 0}v_{g0})v_\parallel-n\Omega_{j}v_{g0}}{\omega_{k0}- k_{\parallel 0}v_{g0}-n\Omega_{j}}.
\end{equation}

The function $\widetilde D_j^n$ in Eq.~(\ref{D}) plays the role of a diffusion coefficient for the $n$ resonance. In $\widetilde D_j^{n}$, the $v_\parallel$-function $W_j^{n}$ defined in Eq.~(\ref{W}) serves as a window function that determines the region in $v_\parallel$-space in which the quasi-linear diffusion through the $n$ resonance is effective. The window function $W_j^{n}$ is maximum at $v_{\parallel res}^n$ defined in Eq.~(\ref{resvelo}), which is the parallel velocity of the particles that resonate with the waves at $k_\parallel=k_{\parallel 0}$ through the $n$ resonance. Our window function $W_j^n$ is linked to Dirac's $\delta$-function in the limit
\begin{equation}\label{delta}
\lim_{v_{\parallel res}^n \rightarrow v_{g0} } W_j^{n} \approx \sqrt{\pi}\frac{\sigma_{\parallel0}}{|k_{\parallel 0}|} \delta(v_\parallel-v_{g0}),
\end{equation}
where $|k_{\parallel 0}|\gg\sigma_{\parallel0}$. Through this ordering between $|k_{\parallel 0}|$ and $\sigma_{\parallel 0}$, we assume that $W_j^{n}$ restricts a finite region in $v_\parallel$-space and that the $W_j^{n}$ for different resonances do not overlap with each other in $v_\parallel$-space.

Only particles distributed within $W_j^{n}$ experience the $n$ resonance and contribute to the quasi-linear diffusion, which is ultimately responsible for the stabilization. Because all terms in Eq.~(\ref{Hn}) are positive semi-definite, all resonances independently stabilize $f_j$ through quasi-linear diffusion in the $v_\parallel$-range defined by their respective $W_j^{n}$, according to Eq.~(\ref{Htt}). Therefore, $H$ decreases and $d H/d t$ tends toward zero during the quasi-linear diffusion through all resonances while $f_j$ is in the process of stabilization. When $f_j$ reaches a state of full stabilization through all $n$ resonances, the instability has saturated and its growth ends.

%Due to $W_j^{n}$ in $D_j^{n}$, all terms for a given $n$ in Eq.~(\ref{Htt}) provide their contributions only in a finite $v_\parallel$-range defined by non-overlapping $W_j^{n}$. 
The $v_\parallel$-function $k_{\parallel j}^{n}$ defined in Eq.~(\ref{res}) is the resonant parallel wavenumber, fulfilling the condition that $k_{\parallel j}^n=k_{\parallel 0}$ at $v_{\parallel}=v_{\parallel res}^n$. It quantifies the $k_\parallel$-component of the unstable $\mathbf{k}$-spectrum in the $v_\parallel$-range defined by $W_j^{n}$. Eq.~(\ref{respv}) defines the phase velocity at $k_{\parallel j}^n$, which is only constant when $v_{g0}=\omega_{k0}/k_{\parallel 0}$, in which case $v_{ph}=v_{g0}$ for all $v_{\parallel}$. We discuss the diffusion operator $\hat{G}[k_{\parallel j}^{n}]$ in Eq.~(\ref{gres}) in the next section.

\subsection{Nature of Quasi-Linear Diffusion in Velocity Space}\label{2.3}

According to Eq.~(\ref{Hn}), unless the wave amplitude is zero, the condition for achieved stabilization through the $n$ resonance is
\begin{equation}\label{con}
    \hat{G}[k_{\parallel j}^{n}] \mathbb{F}_{j}^{n}(v_\perp,v_\parallel)=0, 
\end{equation}
where $\mathbb{F}_{j}^{n}(v_\perp,v_\parallel)$ represents the stabilized VDF of species $j$ through the $n$ resonance. In Eq.~(\ref{con}), $\hat{G}[k_{\parallel j}^{n}]$ is a directional derivative along the isocontour of $\mathbb{F}_{j}^{n}$ evaluated at a given velocity position. Considering the role of $W_j^{n}$, $\hat{G}[k_{\parallel j}^{n}]$ describes only the diffusion of resonant particles within $W_j^{n}$ along the isocontour of $\mathbb{F}_{j}^{n}$. Consequently, the particles experiencing the $n$ resonance diffuse toward the stable state so that $(d H/d t)^n \rightarrow 0$, while the isocontours of $\mathbb{F}_{j}^{n}$ describe the diffusive velocity-space trajectories for the $n$ resonance. 

To find such a trajectory, we express an infinitesimal variation of $\mathbb{F}_{j}^{n}$ along an isocontour as
\begin{equation} \label{fcon}
d\mathbb{F}_{j}^{n}=\frac{\partial \mathbb{F}_{j}^{n}}{\partial v_\perp}d v_\perp+\frac{\partial \mathbb{F}_{j}^{n}}{\partial v_\parallel}d v_\parallel=0.
\end{equation}
Eqs.~(\ref{gres}) and (\ref{fcon}) allow us to rewrite Eq.~(\ref{con}) as
\begin{equation}\label{bc}
    v_\perp d v_\perp \!+\!\left [\frac{n\Omega_{j}}{n\Omega_{j}-\omega_{k0}+ k_{\parallel 0}v_{g0}}\right]\!\!(v_\parallel-v_{g0} ) d v_\parallel=0.
\end{equation}
By integrating Eq.~(\ref{bc}), the diffusive trajectory for the $n$ resonance is then given by
\begin{equation}\label{path}
\begin{split}
    v_\perp^2 \!+\!\left [\frac{n\Omega_{j}}{n\Omega_{j}-\omega_{k0}+ k_{\parallel 0}v_{g0}}\right] \!\!(v_\parallel-v_{g0} )^2=const.
\end{split}
\end{equation}
\citet{1966PhFl....9.2377K} treat the two limiting cases in which $v_{g0}=\omega_{k0}/k_{\parallel 0}$ and $v_{g0}=0$. Using their assumptions, our Eq.~(\ref{path}) is equivalent to their equation (4.8) if $v_{g0}=\omega_{k0}/k_{\parallel 0}$, and our Eq.~(\ref{path}) is equivalent to their equation (4.11) if $v_{g0}=0$. Depending on the dispersion properties of the resonant waves, Eq.~(\ref{path}) is either an elliptic or a hyperbolic equation when $n \neq 0$. In the case of electron resonances, it is safe to assume that
\begin{equation}\label{eta}
    \frac{n\Omega_{j}}{n\Omega_{j}-\omega_{k0}+ k_{\parallel 0}v_{g0}}\geq 0,
\end{equation}
in Eq.~(\ref{path}) if $v_{g0}<(\omega_{k0}+n|\Omega_{e}|)/k_{\parallel 0}$ for all positive $n$ and $v_{g0}>(\omega_{k0}+n|\Omega_{e}|)/k_{\parallel 0}$ for all negative $n$. However, in the case of proton resonances, resonant waves are more likely to violate Eq.~(\ref{eta}) because $\Omega_p\ll |\Omega_e|$.
%In our analysis, we assume that Eq.~(\ref{eta}) is fulfilled. Then, Eq.~(\ref{path}) describes an elliptic diffusion path centered around $\omega_{k0}^\prime$ with major $v_\parallel$ or $v_\perp$-axis, and Eq.~(\ref{path}) with $n=0$ describes an infinite ellipse with major $v_\parallel$-axis.

%Without loss of generality, we set $\omega_{k}>0$. In the case of $k_\parallel>0$, the phase velocity spectrum of $v_{ph}$ in $W_j^{n}$ is in positive values. Then, the coefficient of $\partial/\partial v_\parallel$ at $\hat{G}[{k_{\parallel j}^{n}}]$ is positive in $W_j^{n}$. Considering Eq.~(\ref{eta}), the coefficient of $\partial/\partial v_\perp$ at $\hat{G}[{k_{\parallel j}^{n}}]$ is negative when $v_\parallel> \omega_{k0}^\prime$, but, positive when $v_\parallel<\omega_{k0}^\prime$. Therefore, with $k_\parallel>0$, $\hat{G}[{k_{\parallel j}^{n}}]$ is the directional derivative along Eq.~(\ref{path}) in the clockwise. In the case of $k_\parallel<0$, the phase velocity spectrum of $v_{ph}$ in $W_j^{n}$ is in negative values. $\hat{G}[{k_{\parallel j}^{n}}]$ is then a directional derivative along Eq.~(\ref{path}) in anti-clockwise.

\begin{figure}[ht]
  \centering
  \includegraphics[width=0.47\textwidth]{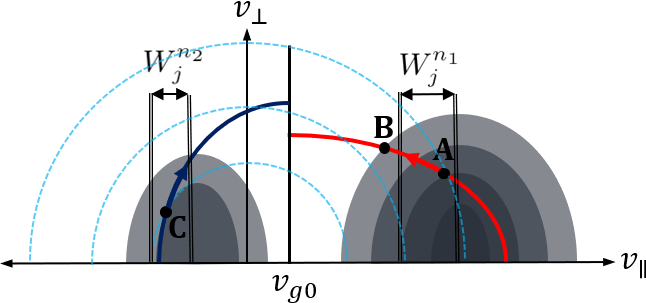}
  \caption{The diffusive flux of resonant particles in velocity space under the action of two arbitrary ($n_1$ and $n_2$) resonances. The dark shaded areas represent isocontours of the VDFs of two particle populations. The red and dark-blue solid curves show the diffusive trajectories, Eq.~(\ref{path}) with $n=n_1$ and $n=n_2$. $W_j^{n_1}$ and $W_j^{n_2}$ represent the window functions according to Eq.~(\ref{W}), in which the $n_1$ and $n_2$ resonances are effective. The light-blue dashed semi-circles correspond to constant-energy contours. The black solid line indicates $v_\parallel=v_{g0}$.}
  \label{exam}
\end{figure}

Fig.~\ref{exam} illustrates the diffusive flux of particles experiencing two arbitrary resonances: the $n_1$ and $n_2$ resonances for an unstable wave. The dark shaded areas represent isocontours of the VDFs of two particle populations in velocity space. The red and dark-blue solid curves represent the diffusive trajectories according to Eq.~(\ref{path}) with $n=n_1$ and $n=n_2$, assuming that the resonant wave fulfills Eq.~(\ref{eta}). The window functions $W_j^{n_1}$ and $W_j^{n_2}$ describe the $v_\parallel$-ranges in which the $n_1$ and $n_2$ resonances are effective. The light-blue dashed semi-circles correspond to contours of constant kinetic energy in the proton rest frame, for which
\begin{equation}\label{cke}
    v_\perp^2+v_\parallel^2=const.
\end{equation} 

In general, the diffusive flux is always directed from higher to lower phase-space densities during the process of stabilization. At point A, resonant particles in $W_j^{n_1}$ diffuse along the red solid curve toward smaller $v_{\parallel}$. Considering the relative alignment between the diffusive flux and the constant-energy contour at point A, the diffusing particles lose kinetic energy. This energy is transferred to the resonant wave, which consequently grows in amplitude. Therefore, this situation corresponds to an instability of the resonant wave. At point B, particles do not diffuse along the red solid curve because this point lies outside $W_j^{n_1}$. %At this point, if $k_\parallel>0$, $\hat{G}[{k_{\parallel j}^{n}}] f_{j}>0$ while if $k_\parallel<0$, $\hat{G}[{k_{\parallel j}^{n}}] f_{j}<0$. 

At point C, resonant particles in $W_j^{n_2}$ diffuse along the dark-blue solid curve toward greater $v_{\parallel}$. Considering the relative alignment between the diffusive flux and the constant-energy contour at point C, the diffusing particles gain kinetic energy. This energy is taken from the resonant wave, which consequently shrinks in amplitude. Therefore, this situation corresponds to damping of the resonant wave and counter-acts the driving of the instability through the $n_1$ resonance. Because the resonant wave is unstable, the $n_1$ resonant instability must overcome the counter-acting $n_2$ resonant damping. %At this point, if $k_\parallel>0$, $\hat{G}[{k_{\parallel j}^{n}}] f_{j}<0$ while if $k_\parallel<0$, $\hat{G}[{k_{\parallel j}^{n}}] f_{j}>0$. Therefore, the sign of $\hat{G}[{k_{\parallel j}^{n}}] f_{j}$ is a discriminant for the diffusive direction along the diffusion path.

According to Eq.~(\ref{Hn}), there are three factors that determine the diffusion rate for the action of an $n$ resonance. The first factor is the particle density $f_j$ within $W_j^n$. The second factor is $\widetilde D_j^n$ whose magnitude is determined by the polarization properties of the resonant waves. The third factor is the quantity $\hat{G}[k_{\parallel j}^{n}]f_j/f_j$ which defines the relative alignment between the isocontours of $f_i$ and the diffusive flux along the diffusion trajectory within $W_j^n$. In Fig.~\ref{exam}, the magnitude of $|\hat{G}[k_{\parallel j}^{n_1}]f_j/f_j|$ at point A is greater than the magnitude of $|\hat{G}[k_{\parallel j}^{n_2}]f_j/f_j|$ at point C. 

Because the diffusive flux is directed from higher to lower values of $f_j$, the quantity $\hat{G}[k_{\parallel j}^{n}]f_j/f_j$ resolves the ambiguity in the directions of the trajectories for resonant particles. A careful analysis of $\hat{G}[k_{\parallel j}^{n}]$ using Eq.~(\ref{eta}) shows that, if $(k_\parallel/|k_\parallel|)(\hat{G}[k_{\parallel j}^{n}]f_j/f_j)>0$ at a given resonant velocity, resonant particles diffuse toward a smaller value of $v_\parallel$ along the diffusive trajectory, while if $(k_\parallel/|k_\parallel|)(\hat{G}[k_{\parallel j}^{n}]f_j/f_j)<0$ at a given resonant velocity, resonant particles diffuse toward a greater value of $v_\parallel$.

\subsection{Numerical Analysis of the Quasi-Linear Diffusion Equation}\label{2.4}

%We numerically solve Eq.~(\ref{QLDQ}) in order to simulate and quantify the quasi-linear velocity-space diffusion.
To simulate the VDF evolution and to compare the diffusion rates between resonances quantitatively, a rigorous numerical analysis of Eq.~(\ref{QLDQ}) is necessary. For this purpose, we develop a mathematical approach based on the Crank-Nicolson scheme and present the mathematical details in Appendix~\ref{A}. Our approach is applicable to all two-dimensional diffusion equations with off-diagonal diffusion terms. Our numerical solution, given by Eq.~(\ref{asol}), evolves the VDF under the action of multiple resonances in one time step. We tested our numerical solution by showing that the diffusive flux obeys the predicted diffusion properties discussed in Section~\ref{2.3}.

\section{Fast-Magnetosonic/Whistler wave and ELECTRON-STRAHL SCATTERING}\label{3}
As an example, we apply our model developed in Section~\ref{2} to an electron resonant instability in the solar wind. The fast-magnetosonic/whistler (FM/W) wave propagating in the anti-sunward direction and with an angle of $\sim60^\circ$ with respect to the background magnetic field scatters the electron strahl \citep{2019ApJ...871L..29V,Verscharen_2019}. Because this prediction is based on linear theory, our quasi-linear framework is appropriate for demonstrating the action of this instability on the electron strahl. 

\subsection{Linear Dispersion Relation}\label{3.1}

To find the characteristics of the unstable oblique FM/W wave, we numerically solve the hot-plasma dispersion relation with the NHDS code \citep{Verscharen_2018}. We use the same plasma parameters as \citet{Verscharen_2019}, which are, notwithstanding the wide range of natural variation,
representative for the average electron parameters in the solar wind \citep{2019ApJS..245...24W}. We assume that the initial plasma consists of isotropic Maxwellian protons, core electrons, and strahl electrons. The subscripts $p$, $e$, $c$ and $s$ indicate protons, electrons, electron core, and electron strahl, respectively. 
\begin{figure}[ht]
  \centering
  \includegraphics[width=0.47\textwidth]{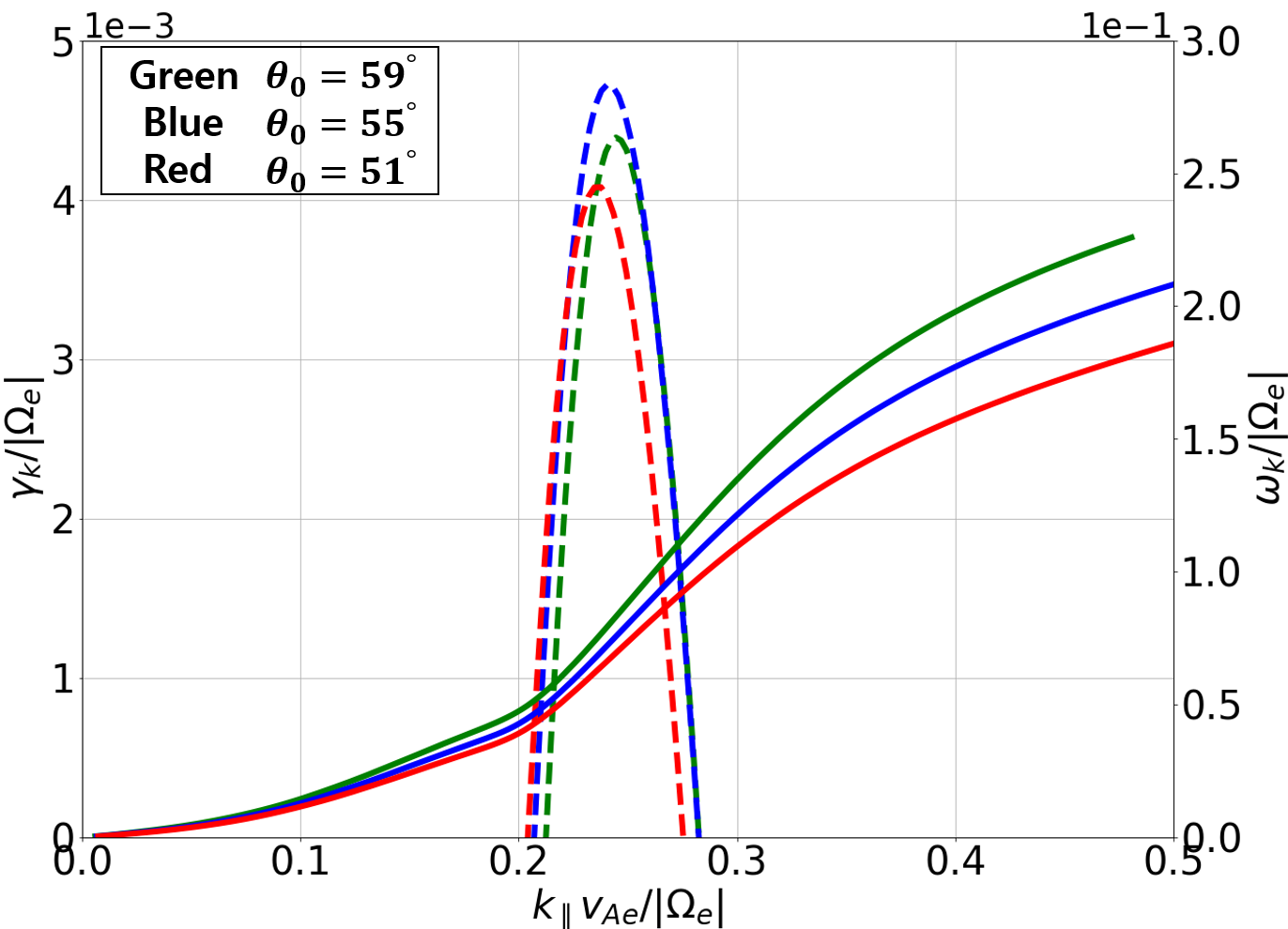}
  \caption{NHDS solutions provide $\gamma_k$ (dashed curves, axis on the left) and $\omega_k$ (solid curves, axis on the right) as functions of the $k_\parallel$-component of the wavevector $\mathbf{k}$. We show solutions for $\theta_0=51^\circ$, $\theta_0=55^\circ$, and $\theta_0=59^\circ$.}
  \label{NHDS}
\end{figure}

We choose our coordinate system so that the anti-sunward and obliquely propagating FM/W waves have $k_{\parallel}>0$. We set $\beta_{c}=\beta_{p}=1$ and $\beta_{s}=0.174$, where $\beta_{j} \! \equiv \! (8\pi n_j k_B T_{j})/B_0^2$, $n_j$ and $T_{j}$ are the density and temperature of species \textit{j}, and $k_B$ is the Boltzmann constant. We set $n_p=n_e$, $n_c=0.92n_p$, $n_s=0.08n_p$, $T_c=T_p$, and $T_s=2T_p$.
In the proton rest frame, we set $n_cU_c+n_sU_s=0$. We initialize the core and strahl bulk velocity with $U_{c}/v_{Ae}\!=-0.22$ and $U_{s}/v_{Ae}\!=2.52$ where $v_{Ae} \! \equiv B_0/\sqrt{4\pi n_e m_e}$ is the electron Alfv\'en speed. NHDS finds that, under these plasma parameters, $\gamma_k>0$ at angles between $\theta_0=51^\circ$ and $\theta_0=67^\circ$. Our strahl bulk velocity then provides a maximum growth rate of $\gamma_k/|\Omega_{e}|=10^{-3}$ \citep{verscharen2019selfinduced}.

Fig.~\ref{NHDS} shows $\gamma_{k}$ and $\omega_{k}$ as functions of the $k_\parallel$-component of the wavevector $\mathbf{k}$ for three different $\theta_0$. The oblique FM/W instability has its maximum growth rate at $\theta_0=55^\circ$, while $\gamma_k>0$ for $0.21\lesssim k_\parallel v_{Ae}/|\Omega_{e}|\lesssim0.28$ which is the parallel unstable $\mathbf{k}$-spectrum.
As defined in Section~\ref{2.1}, we acquire $k_{\parallel 0}v_{Ae}/|\Omega_{e}| \approx 0.245$. This value with Eqs.~(\ref{perk})-(\ref{group}) leads to $k_{\perp 0}v_{Ae}/|\Omega_{e}|=0.35$, $\omega_{k0}/|\Omega_{e}| \approx 0.07$ and $v_{g0}/v_{Ae} \approx 0.86$. We also acquire $\sigma_{\parallel0} v_{Ae}/|\Omega_{e}| \approx 0.035$ and $\sigma_{\perp0}v_{Ae}/|\Omega_{e}|\approx 0.05$ from the unstable $\mathbf{k}$-spectrum.

\subsection{Theoretical Description of the Quasi-Linear Diffusion in the FM/W Instability}\label{3.2}
 
Using the wave and plasma parameters from the previous section, we describe the electron strahl and core diffusion in velocity space. In our analysis, we only consider the $n=+1$, $-1$ and $0$ resonances, ignoring higher-$n$ resonances due to their negligible contributions.

\begin{figure}[ht]
  \centering
  \includegraphics[width=0.45\textwidth]{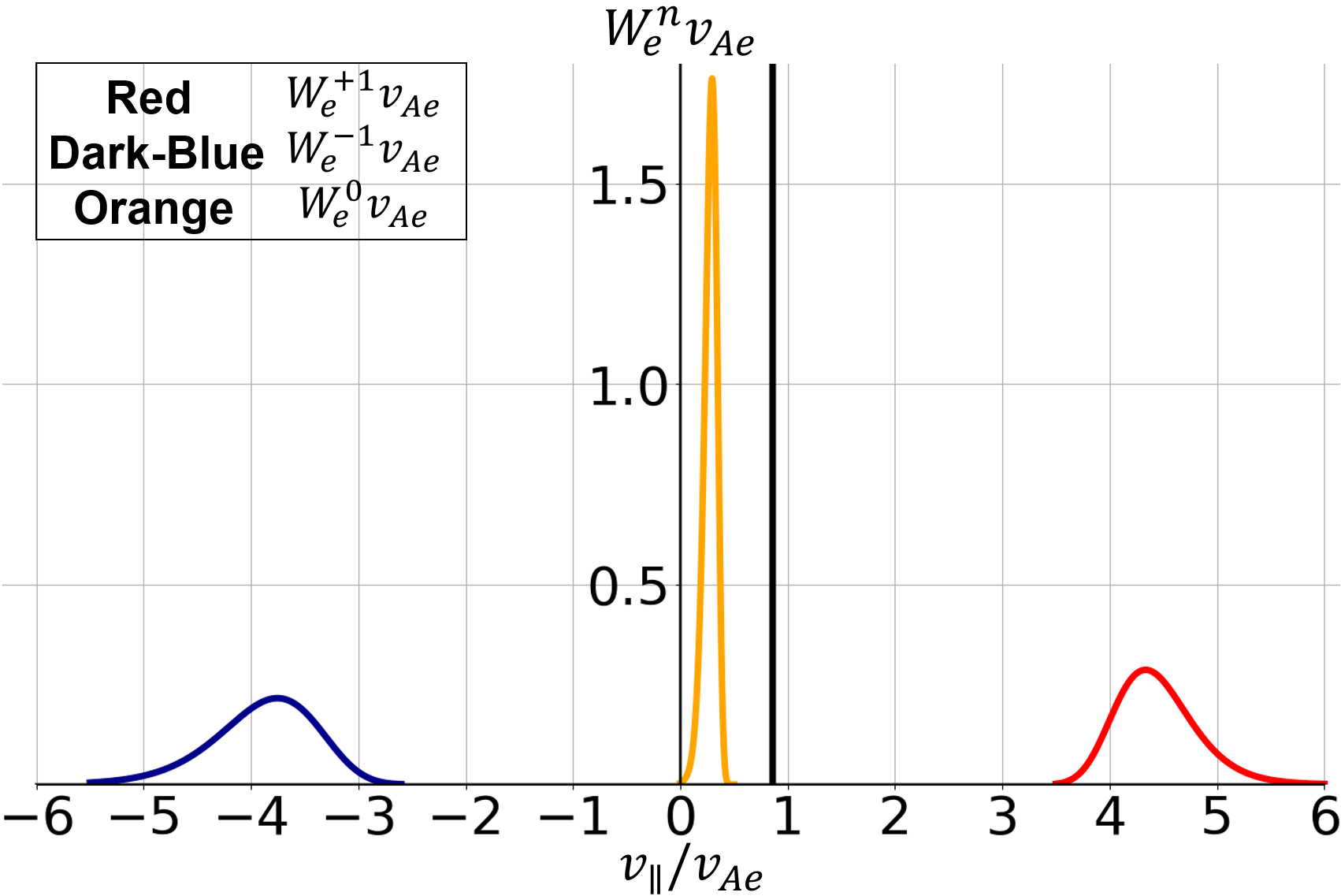}
  \caption{The red, dark-blue, and orange curves illustrate $W_e^{+1}v_{Ae}$, $W_e^{-1}v_{Ae}$ and $W_e^{0}v_{Ae}$ for the oblique FM/W wave. The black solid line represents $v_\parallel=v_{g0}$. Each $W_e^{n}v_{Ae}$ shows the $v_\parallel$-range in which the corresponding resonance is effective. Each $W_e^{n}v_{Ae}$ has a different width in $v_\parallel$-space and maximum value due to a different magnitude of $|v_{\parallel res}^n-v_{g0}|$ (see Eq.~(\ref{delta})).}
  \label{win}
\end{figure}

Upon substituting our wave parameters into Eq.~(\ref{W}), we quantify the dimensionless window functions $W_e^{n}v_{Ae}$ with $n=+1$, $-1$ and $0$. In Fig.~\ref{win}, the red, dark-blue and orange lines represent $W_e^{+1}v_{Ae}$, $W_e^{-1}v_{Ae}$ and $W_e^{0}v_{Ae}$, which are maximum at $v_{\parallel res}^{+1}/v_{Ae}=4.37$, $v_{\parallel res}^{-1}/v_{Ae}=-3.8$ and $v_{\parallel res}^{0}/v_{Ae}=0.29$, respectively. We reiterate that the superscripts indicate the $n$ resonance. The black line indicates $v_\parallel=v_{g0}$. Each $W_e^{n}v_{Ae}$ shows the $v_\parallel$-range in which the quasi-linear diffusion through each resonance is effective. We note that the $W_e^{n}v_{Ae}$ for the three resonances have different widths in $v_\parallel$-space and maximum values due to the different magnitudes of $|v_{\parallel res}^n-v_{g0}|$ (see Eq.~(\ref{delta})). By substituting our wave parameters into Eq.~(\ref{path}), the diffusive trajectories for the $n=+1$, $-1$ and $0$ resonances are given by
\begin{equation}\label{+1}
\left(v_\perp/v_{Ae}\right )^2 +1.16\left (v_\parallel/v_{Ae}-0.86 \right )^2=const,
\end{equation}
\begin{equation}\label{-1}
\left(v_\perp/v_{Ae}\right )^2 +0.88\left (v_\parallel/v_{Ae}-0.86 \right )^2=const,
\end{equation}
and
\begin{equation}\label{0}
\left(v_\perp/v_{Ae}\right )^2 =const.
\end{equation}
Eqs.~(\ref{+1}) and (\ref{-1}) describe ellipses with their axes oriented along the $v_{\perp}$- and $v_{\parallel}$-directions. In Eq.~(\ref{0}), the perpendicular velocity of resonant particles is constant.

\begin{figure}[ht]
  \centering
  \includegraphics[width=0.45\textwidth]{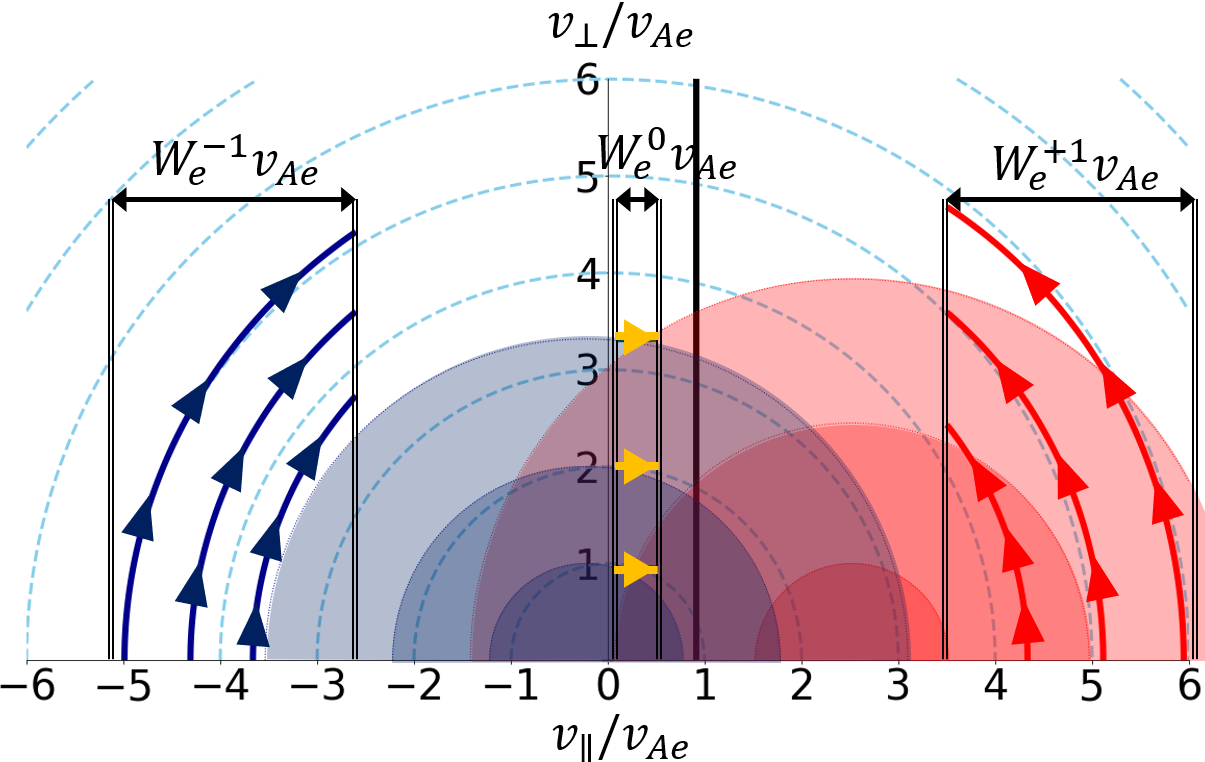}
  \caption{The red, dark-blue, and orange arrows illustrate the diffusive flux for the $n=+1$, $-1$ and $0$ resonances in the oblique FM/W instability. The red and dark-blue filled semi-circles represent isocontours of the strahl and core VDF. This figure does not reflect the relative densities of both electron species. The light-blue dashed semi-circles correspond to constant-energy contours. The black solid line indicates $v_\parallel=v_{g0}$.}
  \label{traject}
\end{figure}

Fig.~\ref{traject} illustrates the electron diffusion from these three resonances. We show the $v_{\parallel}$-ranges in which the three resonances are effective according to $W_e^{+1}v_{Ae}$, $W_e^{-1}v_{Ae}$ and $W_e^{0}v_{Ae}$ from Fig.~\ref{win}. The red, dark-blue, and orange solid curves represent the contours given by Eqs.~(\ref{+1})-(\ref{0}), respectively. The light-blue dashed semi-circles correspond to constant-energy contours in the proton rest frame (see Eq.~(\ref{cke})). The black line indicates $v_\parallel=v_{g0}$. For the initial strahl and core VDF, we apply the plasma parameters in Section~\ref{3.1} to the dimensionless Maxwellian distribution:
\begin{equation}\label{max}
    f_{j}^M=\frac{n_j v_{Ae}^3}{\pi^{3/2}n_p v_{th,j}^3}\exp\left[ -\frac{v_\perp^2+(v_\parallel-U_j)^2}{v_{th,j}^2} \right ],
\end{equation}
where $v_{th,j} \equiv \sqrt{2k_{B}T_{j}/m_j}$. The red and blue areas in Fig.~\ref{traject} represent $f_s^M$ and $f_c^M$ which are normalized by the maximum value of $f_c^M$ and plotted up to a value of $10^{-5}$. In this normalization, Fig.~\ref{traject} does not reflect the relative density ratio between both electron species.

Due to the $v_{\parallel}$-profile of $W_e^{+1}$, the $n=+1$ resonance has a significant effect on $f_{s}^M$. As discussed in Section~\ref{2.3}, because $(k_\parallel/|k_\parallel|)(\hat{G}[k_{\parallel e}^{+1}]f_s^M/f_s^M)>0$, this resonance leads to a diffusion of the resonant strahl electrons in $W_e^{+1}$ along trajectories represented by the red arrows. %\textbf{Since $|dv_\perp/dv_\parallel|$ of Eq.~(\ref{+1}) is smaller than $|dv_\perp/dv_\parallel|$ of Eq.~(\ref{cke}) in $W_e^{+1}$, the distance of resonant strahl electrons from the origin of the coordinate system decreases. This decrease in $v_\perp^2+v_\parallel^2$ represents a loss of kinetic energy of the resonant strahl electrons. The $n=+1$ resonance, therefore, contributes to the driving of the FM/W instability.} 
According to Eq.~(\ref{cke}), the phase-space trajectory of particles that diffuse without a change in kinetic energy is described by
\begin{equation}\label{energpath}
\left( \frac{dv_{\perp}}{dv_{\parallel}}\right)_E=- \frac{v_{\parallel}}{v_{\perp}}.
\end{equation}
 According to Eq.~(\ref{+1}), the phase-space trajectory of resonant particles fulfilling the $n=+1$ resonance, indicated by superscript $+1$, is described by 
\begin{equation}\label{npluspath}
\left(\frac{dv_{\perp}}{dv_{\parallel}}\right)^{+1}=- 1.16\frac{v_{Ae}}{v_{\perp}}\left(\frac{v_{\parallel}}{v_{Ae}}-0.86\right).
\end{equation}
Comparing Eqs.~(\ref{energpath}) and (\ref{npluspath}) in $W_e^{+1}$ shows that $|(dv_{\perp}/dv_{\parallel})^{+1}| < |(dv_{\perp}/dv_{\parallel})_E|$ for the resonant electrons. Therefore, resolving the ambiguity in the directions of the trajectories, the distance of resonant strahl electrons from the origin of the coordinate system decreases. This decrease in $v_{\perp}^2+v_{\parallel}^2$ represents a loss of kinetic energy of the resonant strahl electrons. The $n=+1$ resonance, therefore, contributes to the driving of the FM/W instability.
%Because of $W_e^{+1}$, the diffusion of fast strahl electrons through this resonance is negligible.

Due to the $v_{\parallel}$-profile of $W_e^{-1}$, the $n=-1$ resonance has a significant effect on $f_{c}^M$. Because $(k_\parallel/|k_\parallel|)(\hat{G}[k_{\parallel e}^{-1}]f_c^M/f_c^M)<0$, this resonance leads to a diffusion of the resonant core electrons in $W_e^{-1}$ along trajectories represented by the dark-blue arrows. %\textbf{Since $|dv_\perp/dv_\parallel|$ of Eq.~(\ref{-1}) is larger than $|dv_\perp/dv_\parallel|$ of Eq.~(\ref{cke}) in $W_e^{-1}$, the distance of resonant core electrons from the origin of the coordinate system increases. This increase in $v_\perp^2+v_\parallel^2$ represents a gain of kinetic energy of the resonant core electrons.} 
According to Eq.~(\ref{-1}), the phase-space trajectory of resonant particles fulfilling the $n=-1$ resonance, indicated by superscript $-1$, is described by 
\begin{equation}\label{nminuspath}
\left(\frac{dv_{\perp}}{dv_{\parallel}}\right)^{-1}=- 0.88\frac{v_{Ae}}{v_{\perp}}\left(\frac{v_{\parallel}}{v_{Ae}}-0.86\right).
\end{equation}
Comparing Eqs.~(\ref{energpath}) and (\ref{nminuspath}) in $W_e^{-1}$ shows that $|(dv_{\perp}/dv_{\parallel})^{-1}| > |(dv_{\perp}/dv_{\parallel})_E|$ for the resonant electrons. Therefore, resolving the ambiguity in the directions of the trajectories, the distance of resonant core electrons from the origin of the coordinate system increases. This increase in $v_{\perp}^2+v_{\parallel}^2$ represents a gain of kinetic energy of the resonant core electrons. The $n=-1$ resonance, therefore, counter-acts the FM/W instability through the $n=+1$ resonance.

Due to the $v_{\parallel}$-profile of $W_e^{0}$, the $n=0$ resonance has a significant effect on electrons in the $v_\parallel$-range in which $f_c^M>f_s^M$ and $\partial f_c^M/\partial v_\parallel<0$. Because $(k_\parallel/|k_\parallel|)(\hat{G}[k_{\parallel e}^{0}]f_c^M/f_c^M)<0$, the resonant electrons in $W_e^{0}$ diffuse along trajectories represented by the yellow arrows. Because the distance of these electrons from the origin of the coordinate system increases, these resonant electrons diffuse toward greater kinetic energies. This diffusion removes energy from the resonant FM/W waves and thus counter-acts the driving of the FM/W instability through the $n=+1$ resonance.

Fig.~\ref{traject} only illustrates the nature of the quasi-linear diffusion through the $n=+1$, $-1$ and $0$ resonances in velocity space. It does not give any information regarding the relative strengths of the diffusion rates between the three resonances. Because the FM/W wave is unstable according to linear theory, the $n=+1$ resonant instability must dominate over any counter-acting contributions from the $n=-1$ and $0$ resonances though.

\subsection{Numerical Description of the Quasi-Linear Diffusion in the FM/W Instability}\label{3.3}

We use our numerical procedure from Eq.~(\ref{asol}) to simulate the quasi-linear diffusion through the $n=+1$, $-1$ and $0$ resonances, predicted in Section~\ref{3.2}. According to the definitions in Appendix~\ref{A}, we select the discretization parameters $N_v=60$, $v_{\perp\max}/v_{Ae}=v_{\parallel\max}/v_{Ae}=7$ and $|\Omega_{e}|\Delta t=1$. For the computation of Eq.~(\ref{asol}), we use the same parameters of resonant FM/W waves as those presented in Section~\ref{3.1} and quantify $\widetilde D_e^{\pm 1}$ and $\widetilde D_e^{0}$ in Eq.~(\ref{D}). 

In $\widetilde D_e^{n}$ for each resonance, we only consider the $J_0$ term in $\psi_{j0}^{n}$, ignoring higher-order Bessel functions due to their small contributions. Our NHDS solutions show that $|E_{0}^y| \approx 0.39|E_{0}^x|$ and $|E_{0}^z| \approx 0.28|E_{0}^x|$ in the unstable $\mathbf{k}$-spectrum. Then, we set $|\mathit{{E}}_0^{R}|\approx |\mathit{{E}}_0^{L}| \approx 0.76|\mathit{{E}}_{0}^x|$. Faraday's law yields $\mathit{{E}}_{0}^x \approx [\omega_{k0}/(k_{\parallel 0} c)]\mathit{{B}}_{0}^y$ when ignoring the small contributions from any $E_{0}^z$ terms. This allows us to express $\mathit{{E}}_{0}^x$ through $\mathit{{B}}_{0}^y$ in $\psi_{j0}^{n}$, where $B_{0}^y$ represents the peak amplitude of the wave magnetic-field fluctuations. For simplicity, we assume that $\mathit{{B}}_{0}^y$ is constant in time during the quasi-linear diffusion. Under these assumptions, we acquire
\begin{equation}\label{D+-}
\begin{split}
\widetilde D_e^{\pm 1}&\approx W_e^{\pm 1} \frac{0.58 \pi^2 |\Omega_e|^2 v_{\perp}^2}{ \sigma_{\parallel0} \sigma_{\perp0}^2} \left [\frac{B_{0}^y}{B_0} \frac{\omega_{k0}}{k_{\parallel 0}} \right ]^{2}\\
&\!\!\!\!\times \int_{0}^{\infty} J_0(\rho_e)^2 \exp  \left[-\frac{(k_\perp-k_{\perp0})^2}{\sigma_{\perp0}^2} \right] k_\perp dk_\perp,
\end{split}
\end{equation}
and
\begin{equation}\label{D0}
\begin{split}
\widetilde D_e^{0}&\approx W_e^{0} \frac{0.16 \pi^2 |\Omega_e|^2 v_{\parallel}^2}{ \sigma_{\parallel0} \sigma_{\perp0}^2} \left [\frac{B_{0}^y}{B_0} \frac{\omega_{k0}}{k_{\parallel 0}} \right ]^2 \\
&\!\!\!\!\times \int_{0}^{\infty} J_0(\rho_e)^2 \exp  \left[-\frac{(k_\perp-k_{\perp0})^2}{\sigma_{\perp0}^2} \right] k_\perp dk_\perp,
\end{split}
\end{equation}
where the relative amplitude $B_{0}^y/B_0$ is a free parameter, and we set $B_{0}^y/B_0=0.001$. Then, we apply Eqs.~(\ref{D+-}) and (\ref{D0}) to Eq.~(\ref{asol}).
\begin{figure}[ht]
  \centering
  \includegraphics[width=0.45\textwidth]{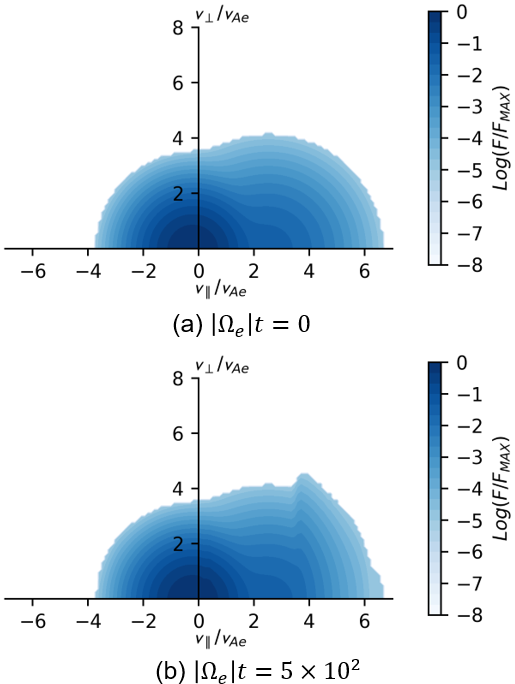}
  \caption{Panel (a): the initial electron VDF; Panel (b): the electron VDF evolved through the $n=+1$, $-1$ and $0$ resonances. Compared to Fig.~\ref{traject}, the effect of the $n=+1$ resonance dominates the evolution during the time $\gamma_kt \sim 1$. It causes a significant pitch-angle gradient at $v_{\parallel}/v_{Ae}\approx 3.8$ through the scattering of strahl electrons. An animation of this figure is available. The animation shows the time evolution of the distribution function from $|\Omega_{e}|t=0$ to $|\Omega_{e}|t=5\times 10^2$. During this evolution, the strahl scattering toward larger $v_{\perp}$ is visible.}
  \label{f5}
\end{figure}

We initialize our numerical computation with the same $f_s^M$ and $f_c^M$ as defined in Section~\ref{3.2}. Fig.~\ref{f5}a represents the normalized $f_e=f_c^M+f_s^M$, plotted up to a value of $10^{-5}$. Fig.~\ref{f5}b shows $f_e$ evolved through the $n=+1$, $-1$ and $0$ resonances, resulting from our iterative calculation of Eq.~(\ref{asol}). Considering the maximum value of the instability's growth rate, $\gamma_k/|\Omega_{e}|=4.8\times10^{-3}$ in Fig.~\ref{NHDS}, we terminate the evaluation of our numerical computation at $|\Omega_{e}|t=5\times10^2$ which corresponds to $\gamma_kt \sim 1$ and thus a reasonable total growth of the unstable FM/W waves.

The strahl electrons at around $v_\parallel/v_{Ae}\approx4.4$ diffuse through the $n=+1$ resonance, as theoretically predicted in Fig.~\ref{traject}. This diffusion increases the pitch-angle of the resonant strahl electrons and generates a strong pitch-angle gradient at $v_\parallel/v_{Ae}\approx3.8$. During this process, the $v_\perp$ of the scattered strahl electrons increases while their $v_{\parallel}$ decreases.  %In addition, we see that strahl electrons at $v_\parallel/v_{Ae} \gtrsim 4.5$ are not scattered and remain at their position in velocity space. 

Because the longitudinal component of the electric-field fluctuations is weaker than their transverse components, the diffusion through the $n=0$ resonance is only slightly noticeable over the modeled time interval. The diffusion through the $n=-1$ resonance is not noticeable even though $\widetilde D_e^{-1}$ and $\widetilde D_e^{+1}$ have similar magnitudes. This is because $|\hat{G}[k_{\parallel e}^{-1}]f_e/f_e|$ in $W_e^{-1}$ is much smaller than $|\hat{G}[k_{\parallel e}^{+1}]f_e/f_e|$ in $W_e^{+1}$, as discussed in Section~\ref{2.3}, and the number of core electrons in $W_e^{-1}$ is very small (see Fig.~\ref{traject} and \ref{f5}).

\section{The secondary effect of Coulomb collisions}\label{4}
Because the collisionless action of resonant wave-particle instabilities often forms strong pitch-angle gradients (see, for example, Fig.~\ref{f5}), collisions can be enhanced in the plasma. Therefore, a more realistic evolution of the total electron VDF must account for the action of Coulomb collisions of strahl electrons with core electrons and protons. For this purpose, we adopt the Fokker-Planck equation given by \citet{doi:10.1098/rspa.1990.0026} with Rosenbluth potentials \citep{PhysRev.107.1} and normalize it in our dimensionless system of units as
\begin{equation}\label{FF}
\begin{split}
\bigg (\frac{\partial  f_{j} }{\partial  t}  \bigg )_{\!\substack{Fokker\\{-Planck}}}
&=\sum_{b} \Gamma_{jb} \bigg \{4\pi \frac{m_j}{m_b} f_b  f_j \\
&\!\!\!\!+ \frac{\partial h}{\partial v^\alpha} \frac{\partial f_j}{\partial v^\alpha}+ \frac{1}{2} \frac{\partial^2 g}{\partial v^\alpha \partial v^\beta}  \frac{\partial^2 f_j}{\partial v^\alpha \partial v^\beta} \bigg \}, 
\end{split}
\end{equation}
where
\begin{equation}
g(\bold v)\equiv\int f_b (\bold v^\prime)|\bold v-\bold v^\prime|d^3 \bold v^\prime,
\end{equation}
\begin{equation}
h(\bold v)\equiv\frac{m_b-m_j}{m_b} \int f_b (\bold v^\prime)|\bold v-\bold v^\prime|^{-1}d^3 \bold v^\prime,
\end{equation}
and
\begin{equation}\label{cp}
\Gamma_{jb}\equiv\frac{4\pi n_b}{v_{Ae}^3 |\Omega_e|}\bigg ( \frac{ Z_j  Z_b q_j^2}{m_j} \bigg )^2  \ln \Lambda_{jb}.
\end{equation}
The subscript \textit{b} indicates the species of background particles, with which the particles of species $j$ Coulomb-collide. The quantity $\ln\Lambda_{jb}$ is the Coulomb logarithm and typically $\ln \Lambda_{jb}\approx 25$ in space plasmas. The parameters $Z_j$ and $Z_b$ are the atomic masses of a particle of species \textit{j} and \textit{b}. The superscripts $\alpha$ and $\beta$ indicate the component of the velocity in cylindrical coordinates and the summation convention holds.

We assume that the timescale of Coulomb collisions is much longer than the timescale of the quasi-linear diffusion in the solar wind under our set of parameters. This assumption allows us to model the resonant wave-particle instability first and to use the resulting VDF as the input for the model of the subsequent, secondary effects of the collisions.
%\begin{equation}\label{QLDFF}
%\frac{\partial  f_{j} }{\partial  t}
%=\bigg (\frac{\partial  f_{j} }{\partial  t}\bigg )_{QLD}+\bigg (\frac{\partial  f_{j} }{\partial  t} \bigg )_{\!\substack{Fokker\\{-Planck}}}.
%\end{equation}

Based on our mathematical approach presented in Appendix~\ref{A}, we present our numerical scheme to solve the Fokker-Planck equation, Eq.~(\ref{FF}), in Appendix~\ref{B}. We tested our numerical solutions, Eq.~(\ref{asolc}), by showing that a set of arbitrary test VDFs diffuses toward $f_b$ with time.

\begin{figure}[ht]
  \centering
  \includegraphics[width=0.45\textwidth]{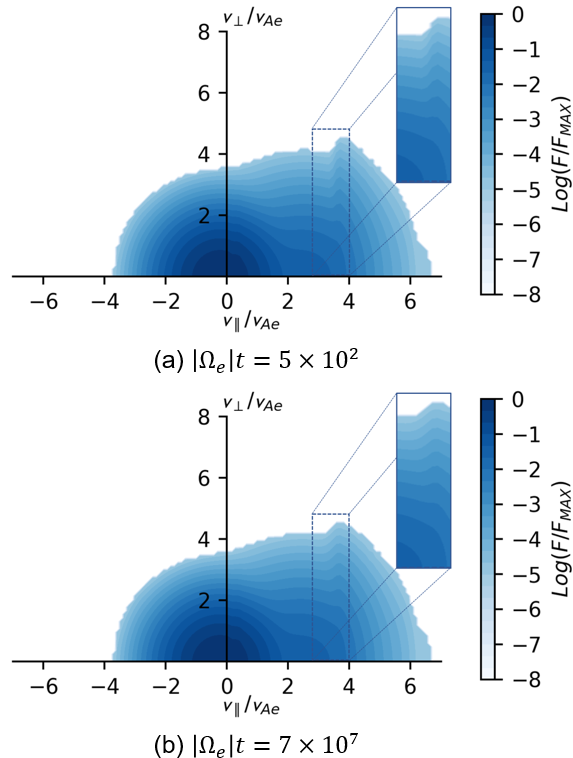}
    \caption{Panel (a): the electron VDF as initial condition for our collision analysis; Panel (b): the electron VDF evolved through Coulomb collisions of strahl electrons with core electrons and protons. The strong pitch-angle gradient at $v_{\parallel}/v_{Ae}\approx 3.8$ (shown in Fig.~\ref{f6}a and Fig.~\ref{f5}b) is relaxed through Coulomb collisions. However, the required time for a noticeable collisional effect on that gradient is around $10^5$ times longer than the timescale of the strahl scattering. An animation of this figure is available. The animation shows the time evolution of the distribution function from $|\Omega_{e}|t=5\times 10^2$ to $|\Omega_{e}|t=7\times 10^7$. During this evolution, the collisional smoothing of the pitch-angle gradients is visible.}
  \label{f6}
\end{figure}

For the computation of Eq.~(\ref{asolc}), we set isotropic Maxwellian electron-core and proton VDFs as background species, $f_b=f_c^M$ and $f_b=f_p^M$, for which we apply the plasma parameters presented in Section~\ref{3.1} to Eq.~(\ref{max}). In this numerical computation, we select the discretization parameters $N_v=60$, $v_{\perp\max}/v_{Ae}=v_{\parallel\max}/v_{Ae}=7$ and $|\Omega_{e}|\Delta t=10$. Moreover, we set $B_0=5\times10^{-4}G$ and $n_{b}=10^2cm^{-3}$ in Eq.~(\ref{cp}), which are representative for the conditions in the solar wind at a distance of 0.3~au from the Sun. We initialize $f_j$ with the electron-strahl VDF $f_s$ from our quasi-linear analysis of the oblique FM/W instability at time $|\Omega_{e}|t=5\times10^2$. In this setup, our initial electron VDF for the Coulomb collision analysis is same as the electron VDF shown in Fig.~\ref{f5}b.

The iterative calculation of Eq.~(\ref{asolc}) results in the time evolution of the electron-strahl VDF under the action of Coulomb collisions with core electrons and protons. The result of this computation at the time $|\Omega_{e}|t=7\times10^7$ is shown in Fig.~\ref{f6}. A detailed comparison of the distribution function before (Fig.~\ref{f6}a) and after (Fig.~\ref{f6}b) our calculation of the effect of Coulomb collisions reveals that Coulomb collisions relax the strong pitch-angle gradient at $v_{\parallel}/v_{Ae}\approx 3.8$, which resulted from the action of the oblique FM/W instability. However, the Coulomb collisions are only capable of affecting strong pitch-angle gradients in the modified electron VDF under our plasma parameters. In addition, the required time for a noticeable collisional effect on this pitch-angle gradient is of order $10^5$ times longer than the characteristic timescale of the quasi-linear diffusion.

\section{Discussion of the strahl scattering}\label{5}
The numerical computation of Eq.~(\ref{Htt}) shows that $dH/dt$ is negative and asymptotically tends toward zero as the electron VDF evolves through the oblique FM/W instability and the counter-acting damping effects until the time $|\Omega_{e}|t=5\times10^2$, which is presented in Fig.~\ref{f5}. Therefore, our quasi-linear diffusion model reflects the stabilization of the particle VDF through the participating wave-particle resonances.

During the action of the oblique FM/W instability, the scattered strahl electrons reduce their collimation along the $\mathbf B_0$-direction and become more isotropic. Even though this instability does not cause significant strahl scattering, we argue that it contributes to the initial formation of the halo population. However, other mechanisms must be considered to account for the full strahl scattering, in agreement with observations \citep{angeo-30-163-2012,angeo-34-1175-2016}.

Alternative models describing Coulomb-collisional effects on the strahl VDF suggest that an anomalous-diffusion process must be considered in order to achieve an agreement with observations \citep{1983JGR....88.6881L,2018MNRAS.474..115H,2019MNRAS.484.2474H}. We note that our analysis includes the subsequent action of Coulomb collisions after the action of collisionless wave-particle resonances assuming plasma parameters consistent with the solar wind at a distance of about 0.3~au from the Sun. Our collisional effects are similar to those proposed by \citet{Vocks_2005}. However, our model predicts that the collisional relaxation is so subtle that the strahl scattering through collisions is barely noticeable for the analyzed phase of the VDF evolution.

The clear separation of timescales between wave-particle effects and  Coulomb-collisional effects complicates the description of the VDF evolution on heliospheric scales, because other processes act on comparable timescales. These additional processes, which our analysis ignores, include turbulence, shocks, plasma mixing, plasma expansion, and magnetic focusing  \citep{doi:10.1029/JA088iA12p09949,doi:10.1029/2003JA009865,doi:10.1063/1.2779282,2012SSRv..173..459Y,article}. A complete model for the radial evolution of the VDF must quantify and account for these processes as well. In the context of our work, these processes can potentially push a VDF that has undergone stabilization as shown in Fig.~\ref{f5}b into the unstable regime again. In this case, $dH/dt$ in Eq.~(\ref{Htt}) returns to a non-zero value, which signifies a new onset of wave-particle resonances and further scattering of resonant particles.
\\
\\
\section{Conclusions}\label{6}
Wave-particle resonances are important plasma-physics processes in many astrophysical plasmas. Often, fully non-linear simulations with codes solving the equations of kinetic plasma theory are used to model the evolution of the distribution function under the action of wave-particle resonances. However, quasi-linear theory augments this approach as it allows us to study the contributions of different processes to these resonances. Therefore, quasi-linear theory is a very helpful tool to improve our understanding of wave-particle resonances in astrophysical plasmas.

We propose a quasi-linear diffusion model for any generalized wave-particle instability. We analyze the quasi-linear diffusion equation by expressing the electric field of an arbitrary unstable and resonant wave mode as a Gaussian wave packet. From Boltzmann's $H$-theorem in our quasi-linear analysis, we define a window function that determines the specific velocity-space range in which a dominant wave-particle instability and counter-acting damping contributions are effective. This window function is the consequence of the localized energy density of our Gaussian wave packet both in configuration space and in wavevector space.

Moreover, we derive a relation describing the diffusive trajectories of the resonant particles for such an instability in velocity space. These trajectories evolve the particle VDF into a stable state in which no further quasi-linear diffusion occurs. Therefore, our theoretical model illustrates the diffusion and stabilization which resonant particles, depending on their location in velocity space, experience in wave-particle resonances.

For the computational quantification of our theoretical model, we introduce a mathematical approach based on the Crank-Nicolson scheme to numerically solve the full quasi-linear diffusion equation. We highlight that this mathematical approach applies to all general two-dimensional diffusion equations, including those with off-diagonal diffusion terms. 

As an example, we apply our model to the oblique FM/W instability that scatters strahl electrons in the solar wind. Our model shows that the $n=+1$ resonant instability of FM/W waves propagating with an angle of $\sim55^\circ$ with respect to the background magnetic field scatters strahl electrons toward larger $v_{\perp}$ and smaller $v_{\parallel}$. The strahl scattering instability through the $n=+1$ resonance dominates over the counter-acting damping contributions through the $n=-1$ and $n=0$ resonances. This instability creates a strong pitch-angle gradient in the electron-strahl VDF.

By numerically solving the Fokker-Planck equation, we show that Coulomb collisions of strahl electrons with core electrons and protons relax this strong pitch-angle gradient on a timescale about $10^5$ times longer than the timescale of the collisionless strahl scattering. This finding suggests that collisional effects are negligible in the strahl-driven oblique FM/W instability, which is a representative example for a resonant wave-particle instability in the solar wind.

Our predicted evolution of the electron VDF is consistent with the observed formation of a proto-halo through strahl scattering \citep{angeo-30-163-2012}. However, further observations are ambiguous regarding the exact source of the proto-halo \citep{angeo-34-1175-2016}. Future high-resolution electron observations with Solar Orbiter and Parker Solar Probe at different distances from the Sun may help us resolve these ambiguities.

Our general quasi-linear diffusion model applies to all non-relativistic collisionless plasmas, such as planetary magnetospheres \citep[e.g.][]{doi:10.1002/2015JA021135}. It also applies to other types of wave-particle instabilities in plasmas such as the resonant instabilities driven by temperature anisotropy or by relative drift. We especially note that our model is also capable of describing ion-driven instabilities.

\acknowledgments
We appreciate helpful discussions with Georgios Nicolaou, Konstantinos Horaites, and Jung Joon Seough. D.V. is supported by the STFC Ernest Rutherford Fellowship ST/P003826/1. D.V., R.T.W., and and A.N.F. are supported by STFC Consolidated Grant ST/S000240/1.

\appendix
\section{Numerical Analysis of the Quasi-linear Diffusion Equation}\label{A}
Eq.~(\ref{QLDQ}) is a second-order differential equation that includes cross-derivative operators such as $\partial^2/\partial v_\parallel \partial v_\perp$. In order to simultaneously evaluate the $\partial^2/\partial v_\parallel \partial v_\perp$ operators
with the $\partial^2/\partial v_\parallel^2$ and $\partial^2/\partial v_\perp^2$ operators in Eq.~(\ref{QLDQ}), we divide velocity space into $2N_v \times 2N_v$ steps with equal step sizes of $\Delta v/2$ by defining the outer boundaries of velocity space as $\pm v_{\perp\max}$ and $\pm v_{\parallel\max}$. The $v_\perp$-index $M$ and the $v_\parallel$-index $N$ both step through 1, 3/2, 2, \dots, $N_v$, $N_v+1/2$. We define the discrete velocity coordinates as $v_{\perp M} \equiv -v_{\perp\max}+(M-1)\Delta v$ and $v_{\parallel N} \equiv -v_{\parallel\max}+(N-1)\Delta v$. We note that this definition introduces negative $v_{\perp}$-values that, although they simplify our numerical analysis, we ignore in our computational results. We divide the time $t$ with equal step sizes of $\Delta t$ and the $t$-index $T$ steps through $1, 2, 3, \cdots$. We define the discrete time as $t^T \equiv (T-1)\Delta t$. We then define the discrete VDF as $f_{M,N}^{T}\equiv f_j( \mathit{v}_{\perp M},\mathit{v}_{\parallel N}, t^T )$. For the discretization of the velocity derivatives, we adopt the two-point central difference operators \citep{Gilat2011NumericalM}
\begin{equation}\label{per}
    \frac{\partial f_j(v_{\perp M},v_{\parallel N},t^T)}{\partial v_{\perp}}\approx \frac{f_{M+1/2,N}^{T}-f_{M-1/2,N}^{T}}{\Delta v},
\end{equation} 
and
\begin{equation}\label{pal}
    \frac{\partial f_j(v_{\perp M},v_{\parallel N},t^T)}{\partial v_{\parallel}}\approx \frac{f_{M,N+1/2}^{T}-f_{M,N-1/2}^{T}}{\Delta v}.
\end{equation}
For the discretization of the time derivative, we adopt the forward difference operator
\begin{equation}\label{time}
    \frac{\partial f_j(v_{\perp M},v_{\parallel N},t^T)}{\partial t}\approx \frac{f_{M,N}^{T+1}-f_{M,N}^{T}}{\Delta t}.
\end{equation} 
By using Eqs.~(\ref{per}) and~(\ref{pal}), we discretize the right-hand side of Eq.~(\ref{QLDQ}) and express it as $(\partial  f /\partial  t)_{M,N}^T$
\begin{equation}\label{dQLD}
\begin{split}
\left (\frac{\partial  f }{\partial  t}\right)_{M,N}^T\equiv  \sum_{n=-\infty }^{\infty } \int \Bigg [ \Bigg(1-\frac{k_\parallel v_{\parallel N}}{\omega_{k0}+v_{g0} (k_\parallel-k_{\parallel 0})} \Bigg)& \frac{1}{v_{\perp M}} \frac{  D_{M+1/2,N}^n [ \hat{G} f ]_{M+1/2,N}^{T} - D_{M-1/2,N}^n [ \hat{G} f  ]_{M-1/2,N}^{T} }{\Delta v}\\
&\!\!\!\!\!\!\!\!\!\!\!\!\!\!\!\!\!\!\!\!\!\!\!\!\!\!\!\!\!\!\!\!\!\!\!\!\!\!\!\!\!\!\!\!\!\!\!\!\!\!\!\!\!\!\!\!\!\!+ \frac{k_\parallel }{\omega_{k0}+v_{g0} (k_\parallel-k_{\parallel 0})} \frac{ D_{M,N+1/2}^n [ \hat{G} f ]_{M,N+1/2}^{T} - D_{M,N-1/2}^n [ \hat{G} f ]_{M,N-1/2}^{T} }{\Delta v} \Bigg ]d^3 \mathbf{k},
\end{split}
\end{equation}
where
\begin{equation}\label{dGT}
[ \hat{G} f  ]_{M,N}^{T} \equiv \!
\Bigg(\!1-\frac{k_\parallel v_{\parallel N}}{\omega_{k0}+v_{g0} (k_\parallel-k_{\parallel 0})} \!\Bigg) \frac{1}{v_{\perp M}} \frac{f_{M+1/2,N}^{T}-f_{M-1/2,N}^{T} }{\Delta v}+ \frac{k_\parallel }{\omega_{k0}+v_{g0} (k_\parallel-k_{\parallel 0})} \frac{f_{M,N+1/2}^{T}-f_{M,N-1/2}^{T} }{\Delta v},
\end{equation}
and
\begin{equation}\label{idD}
\begin{split}
D_{M,N}^n\equiv D_j^{n}\big |_{\substack{v_\perp=v_{\perp M}\\{v_\parallel=v_{\parallel N}}}}.
\end{split}
\end{equation}
According to the Crank-Nicolson scheme \citep{iserles_2008}, the full discretization of Eq.~(\ref{QLDQ}) in its time and velocity derivatives is then given by
\begin{equation}\label{CN}
    f_{M,N}^{T+1}-\frac{\Delta t}{2} \left (\frac{\partial  f }{\partial  t}\right)_{M,N}^
{T+1}=f_{M,N}^{T}+\frac{\Delta t}{2} \left (\frac{\partial  f }{\partial  t}\right)_{M,N}^T.
\end{equation}
By using Eqs.~(\ref{dQLD})-(\ref{idD}) and resolving the $\delta$-functions in $D_{M\pm1/2,N}^n$ and $D_{M,N\pm1/2}^n$ through the $k_\parallel$-integral, we rewrite Eq.~(\ref{CN}) as
\begin{equation}\label{dCN}
\begin{split}
f_{M,N}^{T+1}
-&\sum_{n=-\infty }^{\infty } \Bigg\{\frac{\mu}{2} P_{M,N}^n \widetilde D_{M+1/2,N}^n \left[ P_{M+1/2,N}^n \left (f_{M+1,N}^{T+1}-f_{M,N}^{T+1} \right )+Q_{N}^n \left (f_{M+1/2,N+1/2}^{T+1}-f_{M+1/2,N-1/2}^{T+1} \right ) \right ] \\[2pt]
&- \frac{\mu}{2} P_{M,N}^n \widetilde D_{M-1/2,N}^n \left[ P_{M-1/2,N}^n \left (f_{M,N}^{T+1}-f_{M-1,N}^{T+1} \right )+Q_{N}^n \left (f_{M-1/2,N+1/2}^{T+1}-f_{M-1/2,N-1/2}^{T+1} \right ) \right ] \\[2pt]
&+ \frac{\mu}{2} Q_{N+1/2}^n \widetilde D_{M,N+1/2}^n \left[ P_{M,N+1/2}^n \left (f_{M+1/2,N+1/2}^{T+1}-f_{M-1/2,N+1/2}^{T+1} \right )+Q_{N+1/2}^n \left (f_{M,N+1}^{T+1}-f_{M,N}^{T+1} \right ) \right ] \\[2pt]
&- \frac{\mu}{2} Q_{N-1/2}^n \widetilde D_{M,N-1/2}^n \left[ P_{M,N-1/2}^n \left (f_{M+1/2,N-1/2}^{T+1}-f_{M-1/2,N-1/2}^{T+1} \right )+Q_{N-1/2}^n \left (f_{M,N}^{T+1}-f_{M,N-1}^{T+1} \right ) \right ] \Bigg\}\\[2pt]
=f_{M,N}^{T}
+&\sum_{n=-\infty }^{\infty } \Bigg\{ \frac{\mu}{2} P_{M,N}^n \widetilde D_{M+1/2,N}^n \left[ P_{M+1/2,N}^n \left (f_{M+1,N}^{T}-f_{M,N}^{T} \right )+Q_{N}^n \left (f_{M+1/2,N+1/2}^{T}-f_{M+1/2,N-1/2}^{T} \right ) \right ] \\[2pt]
&- \frac{\mu}{2} P_{M,N}^n \widetilde D_{M-1/2,N}^n \left[ P_{M-1/2,N}^n \left (f_{M,N}^{T}-f_{M-1,N}^{T} \right )+Q_{N}^n \left (f_{M-1/2,N+1/2}^{T}-f_{M-1/2,N-1/2}^{T} \right ) \right ] \\[2pt]
&+ \frac{\mu}{2} Q_{N+1/2}^n \widetilde D_{M,N+1/2}^n \left[ P_{M,N+1/2}^n \left (f_{M+1/2,N+1/2}^{T}-f_{M-1/2,N+1/2}^{T} \right )+Q_{N+1/2}^n \left (f_{M,N+1}^{T}-f_{M,N}^{T} \right ) \right ] \\[2pt]
&- \frac{\mu}{2} Q_{N-1/2}^n \widetilde D_{M,N-1/2}^n \left[ P_{M,N-1/2}^n \left (f_{M+1/2,N-1/2}^{T}-f_{M-1/2,N-1/2}^{T} \right )+Q_{N-1/2}^n \left (f_{M,N}^{T}-f_{M,N-1}^{T} \right ) \right ] \Bigg\},
\end{split}
\end{equation}
where
\begin{equation}\label{p1}
\begin{split}
P_{M,N}^n\equiv \frac{n \Omega_j[v_{\parallel N}-v_{g0}]}{[(\omega_{k0}- k_{\parallel 0} v_{g0}) v_{\parallel N}- n \Omega_j v_{g0}] v_{\perp M} },
\end{split}
\end{equation}
\begin{equation}\label{p2}
\begin{split}
Q_{N}^n\equiv \frac{\omega_{k 0}- k_{\parallel 0}v_{g0} -n\Omega_j }{(\omega_{k0}- k_{\parallel 0} v_{g0}) v_{\parallel N}-n \Omega_j v_{g0}},
\end{split}
\end{equation}
\begin{equation}\label{dD}
\begin{split}
\widetilde D_{M,N}^n\equiv \widetilde D_j^{n}\big |_{\substack{v_\perp=v_{\perp M}\\{v_\parallel=v_{\parallel N}}}},
\end{split}
\end{equation}
and $\mu \equiv \Delta t/\left (\Delta v \right ) ^2$. Eq.~(\ref{dCN}) is a two-dimensional set of algebraic equations, the solution of which, $f_{M,N}^{T+1}$, for all $\mathit{v}_{\perp}$- and $\mathit{v}_{\parallel}$-indexes describes the VDF at time $T+1$ based on $f_{M,N}^T$ for all $\mathit{v}_{\perp}$- and $\mathit{v}_{\parallel}$-indexes.

In order to transform Eq.~(\ref{dCN}) into a single matrix equation with a tridiagonal matrix, we introduce the concept of a double matrix. On both sides of Eq.~(\ref{dCN}), we group the terms by the same $v_\perp$-index in the VDF and rearrange these groups in increasing order in $v_\perp$-index. In each group, we then rearrange terms in increasing order in their $v_\parallel$-index in the VDF. Then, we have 
\begin{equation}\label{rea}
\begin{split}
&\!\!\!\!-\eta ( \mu  )_{M,N}^{(1)} f_{M-1,N}^{T+1}
- \xi ( \mu  )_{M,N}^{(2)} f_{M-1/2,N-1/2}^{T+1}
+ \xi ( \mu  )_{M,N}^{(1)} f_{M-1/2,N+1/2}^{T+1}
-\alpha ( \mu  )_{M,N}^{(2)} f_{M,N-1}^{T+1} 
+ \alpha ( \mu  )_{M,N} f_{M,N}^{T+1}\\[2pt]
&\!\!\!\!-\alpha ( \mu  )_{M,N}^{(1)} f_{M,N+1}^{T+1} 
+ \xi ( \mu  )_{M,N}^{(4)} f_{M+1/2,N-1/2}^{T+1}
- \xi ( \mu  )_{M,N}^{(3)} f_{M+1/2,N+1/2}^{T+1}
-\eta ( \mu  )_{M,N}^{(2)} f_{M+1,N}^{T+1}\\[2pt]
&\!\!\!\!=\\[2pt]
&\!\!\!\!-\eta (\!-\mu \! )_{M,N}^{(1)} f_{M-1,N}^{T}
-\xi (\!-\mu \! )_{M,N}^{(2)} f_{M-1/2,N-1/2}^{T}
+\xi ( \!-\mu \! )_{M,N}^{(1)} f_{M-1/2,N+1/2}^{T}
-\alpha (\!-\mu \! )_{M,N}^{(2)} f_{M,N-1}^{T}+ \alpha ( \!-\mu \! )_{M,N}  f_{M,N}^{T} \\[2pt]
&\!\!\!\!-\alpha ( \!-\mu \! )_{M,N}^{(1)} f_{M,N+1}^{T} 
+\xi ( \!-\mu \! )_{M,N}^{(4)} f_{M+1/2,N-1/2}^{T}
- \xi (\!-\mu \!  )_{M,N}^{(3)} f_{M+1/2,N+1/2}^{T}
-\eta (\!-\mu \! )_{M,N}^{(2)} f_{M+1,N}^{T},
\end{split}
\end{equation}
where
\begin{equation}\label{aph}
\begin{split}
\alpha ( \mu  )_{M,N}
\equiv 1+\frac{\mu}{2} \sum_{n=-\infty }^{\infty } \bigg[ \left (P_{M,N}^n P_{M+1/2,N}^n \right ) \widetilde D_{M+1/2,N}^n & + \left ( P_{M,N}^n P_{M-1/2,N}^n \right ) \widetilde D_{M-1/2,N}^n \\
&\!\!\!+ \left ( Q_{N+1/2}^n \right )^2 \widetilde D_{M,N+1/2}^n+ \left ( Q_{N-1/2}^n \right )^2 \widetilde D_{M,N-1/2}^n\bigg],
\end{split}
\end{equation}
\begin{equation}\label{aph1}
\alpha ( \mu  )_{M,N}^{(1)}\equiv \frac{\mu}{2} \sum_{n=-\infty }^{\infty } \bigg[ \left (Q_{N+1/2}^n \right )^2 \widetilde D_{M,N+1/2}^n \bigg],
\end{equation}
\begin{equation}\label{aph2}
\alpha ( \mu  )_{M,N}^{(2)}\equiv \frac{\mu}{2} \sum_{n=-\infty }^{\infty } \bigg[ \left (Q_{N-1/2}^n \right )^2 \widetilde D_{M,N-1/2}^n \bigg],
\end{equation}
\begin{equation}\label{beta11}
\xi ( \mu  )_{M,N}^{(1)}\equiv \frac{\mu}{2} \sum_{n=-\infty }^{\infty } \bigg[ \left ( P_{M,N}^n Q_N^n \right ) \widetilde D_{M-1/2,N}^n + \left (P_{M,N+1/2}^n Q_{N+1/2}^n \right ) \widetilde D_{M,N+1/2}^n \bigg],
\end{equation}
\begin{equation}\label{beta12}
\xi ( \mu  )_{M,N}^{(2)}\equiv \frac{\mu}{2} \sum_{n=-\infty }^{\infty } \bigg[ \left ( P_{M,N}^n Q_N^n \right ) \widetilde D_{M-1/2,N}^n + \left (P_{M,N-1/2}^n Q_{N-1/2}^n \right ) \widetilde D_{M,N-1/2}^n \bigg],
\end{equation}
\begin{equation}\label{beta21}
\xi(\mu)_{M,N}^{(3)}\equiv \frac{\mu}{2} \sum_{n=-\infty }^{\infty } \bigg[ \left ( P_{M,N}^n Q_N^n \right ) \widetilde D_{M+1/2,N}^n + \left (P_{M,N+1/2}^n Q_{N+1/2}^n \right ) \widetilde D_{M,N+1/2}^n \bigg],
\end{equation}
\begin{equation}\label{beta22}
\xi ( \mu  )_{M,N}^{(4)}\equiv \frac{\mu}{2} \sum_{n=-\infty }^{\infty } \bigg[ \left ( P_{M,N}^n Q_N^n \right ) \widetilde D_{M+1/2,N}^n + \left (P_{M,N-1/2}^n Q_{N-1/2}^n \right ) \widetilde D_{M,N-1/2}^n \bigg],
\end{equation}
\begin{equation}\label{gamma1}
\eta ( \mu  )_{M,N}^{(1)}\equiv \frac{\mu}{2} \sum_{n=-\infty }^{\infty } \bigg[ \left ( P_{M,N}^n P_{M-1/2,N}^n \right ) \widetilde D_{M-1/2,N}^n \bigg],
\end{equation}
and
\begin{equation}\label{gamma2}
\eta ( \mu  )_{M,N}^{(2)}\equiv \frac{\mu}{2} \sum_{n=-\infty }^{\infty } \bigg[ \left (P_{M,N}^n P_{M+1/2,N}^n \right ) \widetilde D_{M+1/2,N}^n \bigg].
\end{equation}
All terms on both sides of Eq.~(\ref{rea}) with a constant $v_{\perp}$-index account for variations in $v_{\parallel}$-space only. Therefore, they can be grouped into a single system of one-dimensional algebraic equations.

We transform all terms with $v_{\perp}$-index $M$ on both sides of Eq.~(\ref{rea}) into the tridiagonal matrices $[A( \mu  )_M][F_M^{T+1}]$ and $[A( -\mu  )_M][F_M^{T}]$, where $F_M^{T}\equiv[
f_{M,1}^{T} \: f_{M,\frac{3}{2}}^{T} \: f_{M,2}^{T} \: \cdots \: f_{M,N_v}^{T}\: f_{M,N_v+\frac{1}{2}}^{T}]^\textbf T_{ 1 \times  2N_v}$ ($\textbf T$ represents the transpose of a matrix), and
\begin{equation}
A( \mu  )_M \!\equiv\!
\begin{bmatrix}
\alpha ( \mu  )_{M,1} \!&\! 0 \!&\! -\alpha ( \mu  )_{M,1}^{(1)} \!&\! 0 \!&\! 0 \!\!&\!\! \cdots \!\!&\!\! 0\\
0 \!&\! \alpha ( \mu  )_{M,3/2} \!&\! 0 \!&\! -\alpha ( \mu  )_{M,3/2}^{(1)} \!&\! 0 \!\!&\!\! \cdots \!\!&\!\! 0\\
-\alpha ( \mu  )_{M,2}^{(2)} \!&\! 0 \!&\! \alpha ( \mu  )_{M,2} \!&\! 0 \!&\! -\alpha ( \mu  )_{M,2}^{(1)} \!\!&\!\! \cdots \!\!&\!\! 0 \\
 \!&\! \vdots \!&\!  \!&\! \ddots \!&\!  \!\!&\!\! \vdots \!\!&\!\!  \\
0 \!&\! \cdots \!&\! 0 \!&\! -\alpha ( \mu  )_{M,N_v}^{(2)} \!&\! 0 \!\!&\!\! \alpha ( \mu  )_{M,N_v} \!\!&\!\! 0 \\
0 \!&\! \cdots \!&\! 0 \!&\! 0 \!&\! -\alpha ( \mu  )_{M,N_v+1/2}^{(2)} \!\!&\!\! 0 \!\!&\!\! \alpha ( \mu  )_{M,N_v+1/2} \\
\end{bmatrix}
_{2N_v \times 2N_v}.
\end{equation}
We transform all terms with $v_{\perp}$-index $M-1/2$ on both sides of Eq.~(\ref{rea}) into the tridiagonal matrices $[B( \mu )_M^{(1)}][F_{M-1/2}^{T+1}]$ and $[B( -\mu  )_M^{(1)}][F_{M-1/2}^{T}]$,
where
\begin{equation}
B( \mu  )_M^{(1)} \equiv
\begin{bmatrix}
0 & \xi ( \mu  )_{M,1}^{(1)} & 0 & \cdots & 0\\
-\xi ( \mu  )_{M,3/2}^{(2)} & 0 & \xi ( \mu  )_{M,3/2}^{(1)}  & \cdots & 0 \\
 & \vdots & \ddots &  & \vdots \\
0 & \cdots  & -\xi ( \mu  )_{M,N_v}^{(2)} & 0 & \xi ( \mu  )_{M,N_v}^{(1)} \\
0 & \cdots  & 0 & -\xi ( \mu  )_{M,N_v+1/2}^{(2)} & 0 \\
\end{bmatrix}
_{2N_v \times 2N_v}.
\end{equation}
We transform all terms with $v_{\perp}$-index $M+1/2$ on both sides of Eq.~(\ref{rea}) into the tridiagonal matrices $[B( \mu  )_M^{(2)}][F_{M+1/2}^{T+1}]$ and $[B( -\mu  )_M^{(2)}][F_{M+1/2}^{T}]$,
where
\begin{equation}
B( \mu  )_M^{(2)} \equiv
\begin{bmatrix}
0 & -\xi ( \mu  )_{M,1}^{(3)} & 0 & \cdots & 0\\
\xi ( \mu  )_{M,3/2}^{(4)} & 0 & -\xi ( \mu  )_{M,3/2}^{(3)}  & \cdots & 0 \\
 & \vdots & \ddots &  & \vdots \\
0 & \cdots  & \xi ( \mu  )_{M,N_v}^{(4)} & 0 & -\xi ( \mu  )_{M,N_v}^{(3)} \\
0 & \cdots  & 0 & \xi ( \mu  )_{M,N_v+1/2}^{(4)} & 0 \\
\end{bmatrix}
_{2N_v \times 2N_v}.
\end{equation}
We transform all terms with $v_{\perp}$-index $M-1$ on both sides of Eq.~(\ref{rea}) into the tridiagonal matrices $[C( \mu  )_M^{(1)}][F_{M-1}^{T+1}]$ and $[C( -\mu  )_M^{(1)}][F_{M-1}^{T}]$,
where
\begin{equation}
C( \mu  )_M^{(1)} \equiv
\begin{bmatrix}
-\eta ( \mu  )_{M,1}^{(1)} & 0  & \cdots & 0\\
0 & -\eta ( \mu  )_{M,3/2}^{(1)}  & \cdots & 0\\
 \vdots &  & \ddots & \vdots \\
0 & \cdots & 0  & -\eta ( \mu  )_{M,N_v+1/2}^{(1)}\\
\end{bmatrix}
_{2N_v \times 2N_v}.
\end{equation}
Lastly, we transform all terms with $v_{\perp}$-index $M+1$ on both sides of Eq.~(\ref{rea}) into the tridiagonal matrices $[C( \mu  )_M^{(2)}][F_{M+1}^{T+1}]$ and $[C( -\mu  )_M^{(2)}][F_{M+1}^{T}]$,
where
\begin{equation}
C( \mu  )_M^{(2)} \equiv
\begin{bmatrix}
-\eta ( \mu  )_{M,1}^{(2)} & 0  & \cdots & 0\\
0 & -\eta ( \mu  )_{M,3/2}^{(2)}  & \cdots & 0\\
 \vdots &  & \ddots & \vdots \\
0 & \cdots & 0  & -\eta ( \mu  )_{M,N_v+1/2}^{(2)}\\
\end{bmatrix}
_{2N_v \times 2N_v}.
\end{equation}
This strategy allows us to express Eq.~(\ref{rea}) as a single system of one-dimensional algebraic equations:
\begin{equation}\label{sol}
\begin{split}
    &[C( \mu  )_M^{(1)}][F_{M-1}^{T+1}]+[B( \mu  )_M^{(1)}][F_{M-1/2}^{T+1}]+[A( \mu  )_M][F_M^{T+1}]+[B( \mu  )_M^{(2)}][F_{M+1/2}^{T+1}]+[C( \mu  )_M^{(2)}][F_{M+1}^{T+1}]\\
    &=[C( -\mu  )_M^{(1)}][F_{M-1}^{T}]+[B( -\mu  )_M^{(1)}][F_{M-1/2}^{T}]+[A\left( -\mu \right )_M][F_M^{T}]+[B(- \mu  )_M^{(2)}][F_{M+1/2}^{T}]+[C(-\mu  )_M^{(2)}][F_{M+1}^{T}].
\end{split}
\end{equation}
Eq.~(\ref{sol}) only describes the VDF evolution in $v_{\perp}$-space. However, each matrix term itself includes the VDF evolution in $v_{\parallel}$-space. We transform Eq.~(\ref{sol}) into a single tridiagonal matrix:
\begin{equation}\label{asol}
E( \mu  )_{QLD}
\begin{bmatrix}
F_1^{T+1} \\
F_{3/2}^{T+1}\\
F_2^{T+1} \\
\vdots \\
F_{N_v+1/2}^{T+1} \\
\end{bmatrix} 
_{\left ( 2N_v \right ) ^2 \times 1}
=
E( -\mu  )_{QLD}
\begin{bmatrix}
F_1^{T} \\
F_{3/2}^{T}\\
F_2^{T} \\
\vdots \\
F_{N_v+1/2}^{T} \\
\end{bmatrix} 
_{\left ( 2N_v \right ) ^2 \times 1},
\end{equation}
where 
\begin{equation}\label{emat}
E( \mu  )_{QLD} \equiv
\begin{bmatrix}
A ( \mu  )_1 & B( \mu  )_1^{(2)} & C( \mu  )_1^{(2)} & 0  & 0 & \cdots & 0\\[4pt]
B( \mu  )_{3/2}^{(1)} & A ( \mu  )_{3/2} & B( \mu  )_{3/2}^{(2)} & C( \mu  )_{3/2}^{(2)} & 0  & \cdots & 0\\[4pt]
C( \mu  )_2^{(1)} & B( \mu  )_2^{(1)} & A \left( \mu \right )_2 & B( \mu  )_2^{(2)} & C( \mu  )_2^{(2)}  & \cdots & 0\\[4pt]
 & \vdots &  & \ddots & & \vdots & \\[4pt]
0 & \cdots  & 0 & C( \mu  )_{N_v}^{(1)} & B( \mu  )_{N_v}^{(1)} & A \left( \mu \right )_{N_v} & B( \mu  )_{N_v}^{(2)}\\[4pt]
0 & \cdots  & 0 & 0 & C( \mu  )_{N_v+1/2}^{(1)} & B( \mu  )_{N_v+1/2}^{(1)} & A \left( \mu \right )_{N_v+1/2}\\
\end{bmatrix}
_{\left ( 2N_v \right ) ^2 \times \left ( 2N_v \right ) ^2}.
\end{equation}
Eq.~(\ref{asol}) is in the form of a double matrix, and $E(\mu)_{QLD}$ in Eq.~(\ref{emat}) defines the evolution matrix. The inner matrices of $E(\mu)_{QLD}$ evolve $f_{M,N}^{T}$ in $v_{\parallel}$-space while the outer matrices of $E(\mu)_{QLD}$ evolve $f_{M,N}^{T}$ in $v_{\perp}$-space during each time step. By multiplying Eq.~(\ref{asol}) with the inverse of $E(\mu)_{QLD}$ on both sides, Eq.~(\ref{asol}) provides the time evolution of $f_{M,N}^T$ in one time step simultaneously in the $v_{\perp}$- and $v_{\parallel}$-spaces. Therefore, it represents the numerical solution of Eq.~(\ref{QLDQ}), which describes the quasi-linear diffusion of a VDF through all resonances.

\section{Numerical Analysis of the Fokker-Planck Equation}\label{B}
In this appendix, we present our numerical strategy to solve the Fokker-Planck equation for Coulomb collisions in Eq.~(\ref{FF}). Using the Crank-Nicolson scheme presented in Appendix~\ref{A}, we discretize Eq.~(\ref{FF}) as
\begin{equation}\label{dFF}
\begin{split}
f_{M,N}^{T+1}
-&\sum_{b}  \frac{\Gamma_{jb}}{2} \bigg[ 4\pi (\Delta v)^2 \mu \frac{m_j}{m_b} f_b(  \mathit{v}_{\perp M},\mathit{v}_{\parallel N}  ) f_{M,N}^{T+1} +\mu  g_{M,N}^{\parallel \perp}  \Big (f_{M+1/2,N+1/2}^{T+1}-f_{M-1/2,N+1/2}^{T+1}-f_{M+1/2,N-1/2}^{T+1}\\
&+f_{M-1/2,N-1/2}^{T+1} \Big )+\frac{\mu  g_{M,N}^{\perp \perp}}{2}  \Big (f_{M+1,N}^{T+1}-2f_{M,N}^{T+1}+f_{M-1,N}^{T+1} \Big )+\frac{\mu g_{M,N}^{\parallel \parallel}}{2}   \Big(f_{M,N+1}^{T+1}-2f_{M,N}^{T+1}+f_{M,N-1}^{T+1} \Big)\\
&+ (\Delta v) \mu h_{M,N}^{\perp}   \Big(f_{M+1/2,N}^{T+1}-f_{M-1/2,N}^{T+1} \Big)+(\Delta v) \mu h_{M,N}^{\parallel}   \Big(f_{M,N+1/2}^{T+1}-f_{M,N-1/2}^{T+1}  \Big) \bigg]\\
=f_{M,N}^{T}
+& \sum_{b} \frac{\Gamma_{jb}}{2} \bigg[ 4\pi (\Delta v)^2 \mu \frac{m_j}{m_b} f_b( \mathit{v}_{\perp M},\mathit{v}_{\parallel N} )  f_{M,N}^{T} +\mu g_{M,N}^{\parallel \perp}  \Big (f_{M+1/2,N+1/2}^{T}-f_{M-1/2,N+1/2}^{T}-f_{M+1/2,N-1/2}^{T}\\
&+f_{M-1/2,N-1/2}^{T} \Big )+\frac{\mu g_{M,N}^{\perp \perp}}{2}   \Big(f_{M+1,N}^{T}-2f_{M,N}^{T}+f_{M-1,N}^{T}  \Big)+\frac{\mu g_{M,N}^{\parallel \parallel}}{2}   \Big(f_{M,N+1}^{T}-2f_{M,N}^{T}+f_{M,N-1}^{T} \Big)\\
&+ (\Delta v) \mu h_{M,N}^{\perp}   \Big(f_{M+1/2,N}^{T}-f_{M-1/2,N}^{T}  \Big)+ (\Delta v) \mu h_{M,N}^{\parallel}   \Big(f_{M,N+1/2}^{T}-f_{M,N-1/2}^{T}  \Big) \bigg],
\end{split}
\end{equation}
where
$g_{M,N}^{ \perp \perp} 
\equiv \partial^2 g/ \partial v_\perp^2$, $g_{M,N}^{\parallel \parallel}
\equiv  \partial^2 g/ \partial v_\parallel^2$, $g_{M,N}^{ \parallel \perp} 
\equiv \partial^2 g/\partial v_\parallel \partial v_\perp$, $h_{M,N}^{\perp} 
\equiv  \partial h/ \partial v_\perp$ and $h_{M,N}^{\parallel} 
\equiv  \partial h/ \partial v_\parallel$, estimated at $v_\perp=v_{\perp M}$ and $v_\parallel=v_{\parallel N}$.

Eq.~(\ref{dFF}) represents a system of two-dimensional algebraic equations. Therefore, we transform Eq.~(\ref{dFF}) into a single tridiagonal matrix using the same strategy for a double matrix as presented in Appendix~\ref{A}.
\begin{equation}\label{asolc}
E( \mu )_{F}
\begin{bmatrix}
F_1^{T+1} \\
F_{3/2}^{T+1}\\
F_2^{T+1} \\
\vdots \\
F_{N_v+1/2}^{T+1} \\
\end{bmatrix} 
_{\left ( 2N_v \right ) ^2 \times 1}
=
E( -\mu  )_{F}
\begin{bmatrix}
F_1^{T} \\
F_{3/2}^{T}\\
F_2^{T} \\
\vdots \\
F_{N_v+1/2}^{T} \\
\end{bmatrix} 
_{\left ( 2N_v \right ) ^2 \times 1},
\end{equation}
where 
\begin{equation}\label{ematc}
E( \mu  )_{F} \equiv
\begin{bmatrix}
X ( \mu )_1 & -Y ( \mu  )_1 & Z ( \mu  )_1 & 0  & 0 & \cdots & 0\\
Y ( \mu  )_{3/2} & X ( \mu  )_{3/2} & -Y ( \mu  )_{3/2} & Z ( \mu  )_{3/2} & 0  & \cdots & 0\\
Z ( \mu  )_2 & Y ( \mu  )_2 & X ( \mu  )_2 & -Y ( \mu  )_2 & Z ( \mu  )_2  & \cdots & 0\\
 & \vdots &  & \ddots & & \vdots & \\
0 & \cdots  & 0 & Z ( \mu  )_{N_v} & Y ( \mu  )_{N_v} & X ( \mu  )_{N_v} & -Y ( \mu  )_{N_v}\\
0 & \cdots  & 0 & 0 & Z ( \mu  )_{N_v+1/2} & Y ( \mu  )_{N_v+1/2} & X ( \mu  )_{N_v+1/2}\\
\end{bmatrix}
_{ ( 2N_v  ) ^2 \times  ( 2N_v  ) ^2},
\end{equation}

\begin{equation}
X( \mu )_M \!\!\equiv\!\!
\begin{bmatrix}
\!\varepsilon ( \mu  )_{M,1} \!&\! -\varepsilon ( \mu  )_{M,1}^{(2)} \!&\! -\varepsilon ( \mu  )_{M,1}^{(1)} \!&\! 0 \!&\! 0 \!&\! \cdots \!&\! 0\!\!\\[4pt]
\!\varepsilon ( \mu  )_{M,3/2}^{(2)} \!&\!\varepsilon ( \mu  )_{M,3/2} \!&\! -\varepsilon ( \mu  )_{M,3/2}^{(2)} \!&\! -\varepsilon ( \mu  )_{M,3/2}^{(1)} \!&\! 0 \!&\! \cdots \!&\! 0\!\!\\[4pt]
\!-\varepsilon ( \mu  )_{M,2}^{(1)} \!&\! \varepsilon ( \mu  )_{M,2}^{(2)} \!&\! \varepsilon ( \mu  )_{M,2} \!&\! -\varepsilon ( \mu  )_{M,2}^{(2)} \!&\! -\varepsilon ( \mu  )_{M,2}^{(1)} \!&\! \cdots \!&\! 0 \!\!\\[4pt]
\!&\!\vdots\!&\!  \!&\!\ddots \!&\!  \!&\! \vdots \!&\!\!  \\[4pt]
\!0 \!&\! \cdots \!&\! 0 \!&\! -\varepsilon ( \mu  )_{M,N_v}^{(1)} \!&\! \varepsilon ( \mu  )_{M,N_v}^{(2)} \!&\! \varepsilon ( \mu  )_{M,N_v} \!&\! -\varepsilon ( \mu  )_{M,N_v}^{(2)} \!\!\\[4pt]
\!0 \!&\! \cdots \!&\! 0 \!&\! 0 \!&\! -\varepsilon ( \mu  )_{M,N_v+1/2}^{(1)} \!&\! \varepsilon ( \mu  )_{M,N_v+1/2}^{(2)} \!&\!\varepsilon ( \mu  )_{M,N_v+1/2}\!\! \\
\end{bmatrix}
_{2N_v \times 2N_v},
\end{equation}
\begin{equation}
Y( \mu  )_M \equiv
\begin{bmatrix}
\varrho ( \mu  )_{M,1}^{(2)} & \varrho ( \mu  )_{M,1}^{(1)} & 0 & 0 & \cdots & 0\\[4pt]
-\varrho ( \mu  )_{M,3/2}^{(1)} & \varrho ( \mu  )_{M,3/2}^{(2)} & \varrho ( \mu  )_{M,3/2}^{(1)} & 0 & \cdots & 0 \\[4pt]
 & \vdots & \ddots & & & \vdots \\[4pt]
0 & \cdots & 0  & -\varrho ( \mu  )_{M,N_v}^{(1)} & \varrho ( \mu  )_{M,N_v}^{(2)} & \varrho ( \mu  )_{M,N_v}^{(1)} \\[4pt]
0 & \cdots  & 0 & 0& -\varrho ( \mu  )_{M,N_v+1/2}^{(1)} & \varrho ( \mu  )_{M,N_v+1/2}^{(2)} \\
\end{bmatrix}
_{2N_v \times 2N_v},
\end{equation}
\begin{equation}
Z( \mu  )_M \equiv
\begin{bmatrix}
-\tau ( \mu  )_{M,1} & 0  & \cdots & 0\\
0 & -\tau ( \mu  )_{M,3/2}  & \cdots & 0\\
 \vdots &  & \ddots & \vdots \\
0 & \cdots & 0  & -\tau ( \mu  )_{M,N_v+1/2}\\
\end{bmatrix}
_{2N_v \times 2N_v},
\end{equation}

\begin{equation}\label{aphc}
\begin{split}
\varepsilon ( \mu  )_{M,N}
\equiv 1-\mu \sum_{b} \Gamma_{jb} \bigg[ 2\pi(\Delta v)^2 \frac{m_j}{m_b}f_b( \mathit{v}_{\perp M}, \mathit{v}_{\parallel N} )-\frac{g_{M,N}^{ \perp \perp} }{2} -\frac{g_{M,N}^{ \parallel \parallel} }{2} \bigg],
\end{split}
\end{equation}
\begin{equation}\label{aph1c}
\varepsilon ( \mu  )_{M,N}^{(1)}\equiv \mu \sum_{b}\frac{\Gamma_{jb}g_{M,N}^{ \parallel \parallel}}{4},
\end{equation}
\begin{equation}\label{aph2c}
\varepsilon ( \mu  )_{M,N}^{(2)}\equiv \mu \sum_{b}\frac{\Gamma_{jb}h_{M,N}^{ \parallel} (\Delta v)}{2},
\end{equation}
\begin{equation}\label{beta11c}
\varrho ( \mu  )_{M,N}^{(1)}\equiv \mu \sum_{b}\frac{\Gamma_{jb}g_{M,N}^{ \parallel \perp}}{2},
\end{equation}
\begin{equation}\label{beta12c}
\varrho( \mu  )_{M,N}^{(2)}\equiv \mu \sum_{b}\frac{\Gamma_{jb}h_{M,N}^{ \perp} (\Delta v)}{2},
\end{equation}
and
\begin{equation}\label{gammac}
\tau ( \mu  )_{M,N}\equiv \mu \sum_{b}\frac{\Gamma_{jb}g_{M,N}^{ \perp \perp}}{4}.
\end{equation}
Like Eq.~(\ref{asol}), Eq.~(\ref{asolc}) provides the time evolution of $f_{M,N}^T$ in one time step simultaneously in the $v_{\perp}$- and $v_{\parallel}$-spaces. Therefore, it represents the numerical solution of Eq.~(\ref{FF}), which describes the action of Coulomb collisions of particles in $f_j$ with particles in $f_b$.
\clearpage


\bibliography{sample63}{}
\bibliographystyle{aasjournal}



\end{document}